\documentclass[twocolumn,,amsmath,amssymb]{revtex4-1}

\pdfoutput=1

\usepackage{graphicx}
\usepackage{dcolumn}
\usepackage{bm}
\newcommand{\etal}{\emph{et al.}}
\newcommand{\be}{\begin{equation}}
\newcommand{\ee}{\end{equation}}
\newcommand{\bfig}{\begin{figure}}
\newcommand{\efig}{\end{figure}}

\usepackage{lineno}
\usepackage{lipsum}

\begin{document}
\title{Review of experiments on the chiral anomaly in Dirac-Weyl semimetals
}
\author{N. P. Ong and Sihang Liang}
\affiliation{
Department of Physics, Princeton University, Princeton, NJ 08544
}

\date{\today}
\begin{abstract}
We provide a review of recent experimental results on the chiral anomaly in Dirac/Weyl semimetals. After a brief introduction, we trace the steps leading to the prediction of materials that feature protected 3D bulk Dirac nodes. The chiral anomaly is presented in terms of charge pumping between the chiral Landau levels of Weyl fermions in parallel electric and magnetic fields. The related chiral magnetic effect and chiral zero sound are described. Current jetting effects, which present major complications in experiments on the longitudinal magnetoresistance, are carefully analyzed. We describe a recent test that is capable of distinguishing these semiclassical artifacts from intrinsic quantum effects. Turning to experiments, we review critically the longitudnial magnetoresistance experiments in the Dirac/Weyl semimetals Na$_3$Bi, GdPtBi, ZrTe$_5$ and TaAs. Alternate approaches to the chiral anomaly, including experiments on non-local transport, thermopower, thermal conductivity and optical pump-probe response are reviewed. In the Supplement, we provide a brief discussion of the chiral anomaly in the broader context of high energy physics and relativistic quantum field theory, as well the anomaly's starring role at the nexus of quantum physics and differential geometry.
\end{abstract}

\maketitle


\section{Introduction}\label{intro}
In the past 2 decades, research on Dirac electrons in semimetals has grown rapidly into a major field of activity, especially in topological quantum matter. Initially, the focus was on the protected two-dimensional (2D) Dirac states in graphene and on the surfaces of topological insulators (TIs). In 2011-2012, the publication of Refs.~\cite{Vishwanath,Ran,Burkov,XiDai,Young} caused a sea change that shifted research from TIs towards Dirac/Weyl semimetals. The search culminated in the discovery of the Dirac semimetals, Na$_3$Bi \cite{WangNa3Bi} and Cd$_3$As$_2$, followed by the Weyl semimetals TaAs and NbAs (Sec. \ref{symmetry}). The realization of 3D Dirac states in semimetals revived intense interest in the chiral anomaly~\cite{Adler,Bell} and its observability in crystals~\cite{Nielsen,Aji,SonSpivak,Sid,Burkov15}.

In quantum field theory (QFT), an anomaly~\cite{Peskin,Nakahara,Bertlmann} is the breaking of a classically allowed symmetry when quantum effects are turned on (see Supplementary Information Sec. \ref{QFT}). The chiral (or axial) anomaly -- the first example discovered~\cite{Adler,Bell} -- remains the most investigated experimentally. In the massless limit, the Lagrangian of Dirac fermions splits into two independent parts describing right-handed and left-handed massless fermions. Conservation of chiral symmetry implies that the two populations $N_R$ and $N_L$ are separately conserved. However, coupling of the fermions to a vector gauge field (the electromagnetic field) breaks the chiral symmetry, resulting in the appearance of the anomaly term. In 1968 the difficulty of calculating the decay rate of the neutral pion $\pi^0$ was resolved by the discovery of the triangular anomaly diagram by Adler~\cite{Adler} and Bell and Jackiw~\cite{Bell}. Subsequently, anomalies have appeared in diverse phenomena occuring at vastly different energy scales (see Supplementary Information for details).  In 1983, Nielsen and Ninomiya~\cite{Nielsen} proposed that the chiral anomaly should be observable in crystals as a large, negative longitudinal magnetoresistance (LMR).

The discovery of Dirac and Weyl semimetals led to a surge of experiments investigating the LMR (Sec. \ref{LMR}). The findings fall into two groups with distinct features. In the semimetals Na$_3$Bi and GdPtBi, which have low carrier mobilities, the negative LMR is very large (25 to 500$\%$) and monotonic, settling down to a constant in high magnetic field $\bf B$. In the second group, comprised of the Weyl semimetals TaAs and NbAs, which have very high mobilities, the LMR is strongly non-monotonic; it is weakly negative (0.1 to 2$\%$) in small $B$ but becomes positive above 1 T. Inconsistencies soon raised strong concerns that the observed LMR in the second group is an artifact caused by $B$-induceed inhomogeneous current flow known as current jetting (Sec. \ref{jetting}). Recently, Liang \etal ~\cite{Liang} proposed a test that distinguishes the intrinsic LMR from current jetting artifacts. The test confirms that the negative LMR is intrinsic in the first group but artifactual in the second (closely similar to LMR results in bulk Bi and InAs). We review the results in Sec. \ref{LMR}, as well as other experimental approaches to the chiral anomaly (Sec. \ref{related}). A recent review of Dirac semimetals complementary to our review is Ref.~\cite{Armitage}.

The acronyms used are ABJ (Adler Bell Jackiw) ARPES (angle-resolved photoemission spectroscopy), CME (chiral magnetic effect), CVD (chemical vapor deposition), CZS (chiral zero sound), IS (inversion symmetry), LL0 (Landau level with $n = 0$), LMR (longitudinal magnetoresistance), QCD (quantum chromodynamics), QFT (quantum field theory), SdH (Shubnikov de Haas), TI (topological insulator), TRI (time-reversal invariance), and TRIM (time-reversal invariant momenta).

\section{Symmetry Protection and Weyl Nodes}\label{symmetry}
Initially, the search for materials with protected 3D Dirac and Weyl states~\cite{Vishwanath,Ran,Burkov,XiDai,Young} focussed on the two symmetries, time-reversal invariance (TRI) and inversion symmetry (IS). These are sufficient to protect 3D Dirac nodes that are located at the time-reversal invariant momenta (TRIM) $\bf K$ (which satisfy $\bf K = -K +G$ with $\bf G$ a reciprocal lattice vector). The predicted materials, notably $\beta$-crystobalite BiO$_2$~\cite{Young}, are unfortunately unstable. 

[To illustrate protection by TRI at a TRIM $\bf K$, we take $\Psi_{\bf K}$ and $\Phi_{\bf K}$ to be 2-spinor eigenstates at $\bf K$ that are time-reversed partners. Then by the Kramers theorem, they must be orthogonal, i.e. $\langle \Psi_{\bf K}, \Phi_{\bf K}\rangle$ = 0. If $V$ is a weak potential that is invariant under TR, the matrix element $\langle \Psi_{\bf K}, V\Phi_{\bf K}\rangle$ also vanishes by the same argument. Hence TRI protects the node at $\bf K$ against gap formation induced by $V$.]

Subsequently, it was realized that including point-group symmetry confers protection of Dirac nodes anywhere along a symmetry axis (e.g. the rotation axis of ${\cal C}_n$, with $n=3,4,6$). Freeing the node from a TRIM led to the prediction (Wang \etal~\cite{WangNa3Bi}) that the two semimetals Na$_3$Bi~\cite{WangNa3Bi} and Cd$_3$As$_2$ are Dirac semimetals protected by symmetry under ${\cal C}_3$ and ${\cal C}_4$ rotations, respectively.

In the case of Na$_3$Bi (Sec. \ref{Na3Bi}), the workhorse for the chiral anomaly, Wang \etal~ showed that, with the choice of basis for the 4-spinor
\be
\hat{\Psi} = ( \;|S,+\rangle, |P, +\rangle, |S,-\rangle, |P, -\rangle \;)^T
\label{4spinor}
\ee
(where $[\cdots]^T$ indicates transpose, and $\pm$ refer to the $z$ component of $J_z$), the linearized Hamiltonian close to the Dirac node at ${\bf k}_{D+}$ splits into two $2\times 2$ blocks given by
\be
H({\bf q}) = v_F\left(
		\begin{array}{cccc}
		\beta q_z & q_+ & 0 & 0 \\
		q_- & -\beta q_z  &  0 & 0 \\
		0  & 0   &  \beta q_z & -q_- \\
		0  & 0   &  -q_+ & -\beta q_z
		\end{array}
		\right),
\label{Hmatrix}
\ee
where $\beta<1$ is the velocity anisotropy and ${\bf q = k-k}_{D+}$.

The upper and lower blocks are compactly written as 
\begin{eqnarray}
H_-  &=&  v_F( q_x\tau_x - q_y\tau_y + \beta q_z\tau_z), \nonumber \\ 
H_+ &=&  v_F( -q_x\tau_x - q_y\tau_y + \beta q_z\tau_z),
\label{HWeyl}
\end{eqnarray}
where $\{\tau_i \}$ are Pauli matrices acting in the orbital subspace $(S,P)$.

$H_{\mp}$ represent Weyl nodes of chirality $\chi=\mp$ ($\chi$ is the determinant of the velocity matrix $\tilde{V}_{\pm}/v_F$ defined by the relation $H_{\pm} = {\bf q}\cdot \tilde{V}_{\pm} \cdot \boldsymbol{\tau}$.) 

In zero magnetic field, the two Weyl nodes coincide in $\bf k$ space. If TRI is broken in finite $\bf B$, the Zeeman energy leads to their separation. A characteristic of Weyl states is that the $n=0$ Landau level is chiral. As discussed in Sec. \ref{Landau}, in an applied $\bf E$, the chiral anomaly is predicted for carriers occupying the $n =0$ level.

A detailed symmetry analysis of space group symmetries that can protect 3D Dirac nodes has been reported by Yang and Nagaosa~\cite{YangNagaosa}. With the choice for the inversion operator ${\cal P} = \tau_z$, they show how the distinct eigenvalues of the point-group rotation ${\cal C}_n$ protect the Dirac nodes for $n= 3$ (Na$_3$Bi) and $n=4$ (Cd$_3$As$_2$). The distinct values of the eigenvalues implies that all matrix elements formed between states in the conduction and valence bands must vanish. If ${\cal P} = \tau_x$ (the case of $\beta$-BiO$_2$), the Dirac node is pinned at a TRIM. The remaining case ${\cal P} = \boldsymbol{1}$ has not yet led to a candidate.

\section{Landau quantization and chiral anomaly}
\subsection{Chiral Anomaly}\label{Landau}
Following the publication of Ref. \cite{Vishwanath}, several approaches for detecting the chiral anomaly were proposed~\cite{Burkov,Aji,SonSpivak,Sid,Burkov15}. We review the approach that exploits the chiral nature of the lowest Landau level in 3D bulk Weyl fermions.

In a strong magnetic field $\bf B\parallel \hat{z}$, the bulk electronic states in a conventional semimetal are quantized into Landau levels (LL) indexed by $n = 0, 1, 2, \cdots$. For each of the LLs, the energy disperses as $E(n,k_z) = (\hbar k_z)^2/2m_z$ where $m_z$ is the effective mass for dispersion $\parallel\bf\hat{z}$.

A distinguishing feature of Weyl nodes is that the LL with $n=0$ (hereafter called LL0) is chiral, i.e. $E(0, k_z) = \pm v_Fk_z$ (Fig. \ref{figLandau}a). For the Weyl node with $\chi$ = +1, the slope is positive, i.e. the velocity $\bf v\parallel B$, whereas for $\chi = -1$, $\bf v\parallel -B$. Hence, when $B$ exceeds $B_Q$ (the field at which the chemical potential $\mu$ enters the LL0), we have two independent populations $N_L$ and $N_R$ of massless left-and right-moving massless fermions. Because they do not intermix, their respective current densities ${\bf J}_R$ and ${\bf J}_L$ are independently conserved. Hence the total current density ${\bf J} = {\bf J}_L + {\bf J}_R$ and difference current density ${\bf J}^5 = {\bf J}_L-{\bf J}_R$ are also conserved, viz.
\be
\partial_t \rho + \nabla\cdot {\bf J} = 0, \quad  \partial_t \rho^5 + \nabla\cdot {\bf J}^5 = 0,
\label{conserve}
\ee
where $\rho$ and $\rho^5$ are the respective charge densities (superscripts $^5$ reflect the chirality matrix $\gamma_5$ (Sec. \ref{QFT})). This represents chiral symmetry of massless fermions in the LL0.

Application of an electric field $\bf E\parallel B\parallel \hat{z}$ causes the population of the branch moving down the potential incline (say $N_R$) to increase while the uphill branch decreases at the same rate -- the chiral symmetry is broken by coupling to $\bf E$ and $\bf B$. The ``pumping rate'' $dN^5/dt$ is the product of the 2D density of states ${\cal D}_0 = L^2/(2\pi\ell_B^2)$ of the LL0 and the rate of increase of available states driven by $\bf E$, $dn/dt = (L/2\pi)dk_z/dt$ (where $L^3$ is the sample volume and $\ell_B= \sqrt{\hbar/eB}$ the magnetic length). We have
\be
\frac{1}{L^3}\frac{dN^5}{dt} = \frac{1}{2\pi\ell_B^2}\frac{1}{2\pi}\frac{eE}{\hbar} = \frac{e^2}{4\pi^2\hbar^2} {\bf E\cdot B} \equiv {\cal A}.
\label{AA}
\ee

The anomaly term ${\cal A}$ acts like a source term that ruins conservation of the current ${\bf J}^5$, which we express as
\be
\partial_t \rho^5 + \nabla\cdot {\bf J}^5 = {\cal A},
\label{J5cons}
\ee
where $\rho^5 = N^5/L^3$. Equation \ref{J5cons}, which represents the chiral anomaly in a semimetal, implies the emergence of a new axial charge current that strongly enhances the conductivity $\sigma_{zz}$ if the axial relaxation time $\tau_A$ greatly exceeds $\tau_0$, the conventional transport lifetime. The enhancement is observable as a large, negative LMR ($\bf B\parallel E$).

\subsection{Chiral Magnetic Effect and Zero Sound} \label{CME}
Although they share a common origin, the chiral anomaly is distinct from the ``chiral magnetic effect'' (CME)~\cite{Kharzeev,ChenBurkov} which predicts that a magnetic field $\bf B$ applied to a Weyl semimetal (with $\bf E =0$) spontaneously generates an axial charge current $\parallel\bf B$ or $-\bf B$, depending on the chirality $\chi$. In condensed matter, the appearance of a DC (transport) current driven by an applied $\bf B$ acting alone violates the laws of thermodynamics. Franz and collaborators~\cite{Franz} have shown that the CME vanishes in calculations that are properly regularized. The CME by itself cannot exist in magnetoresistance experiments.

\begin{figure}
\centering
\includegraphics[width=14cm]{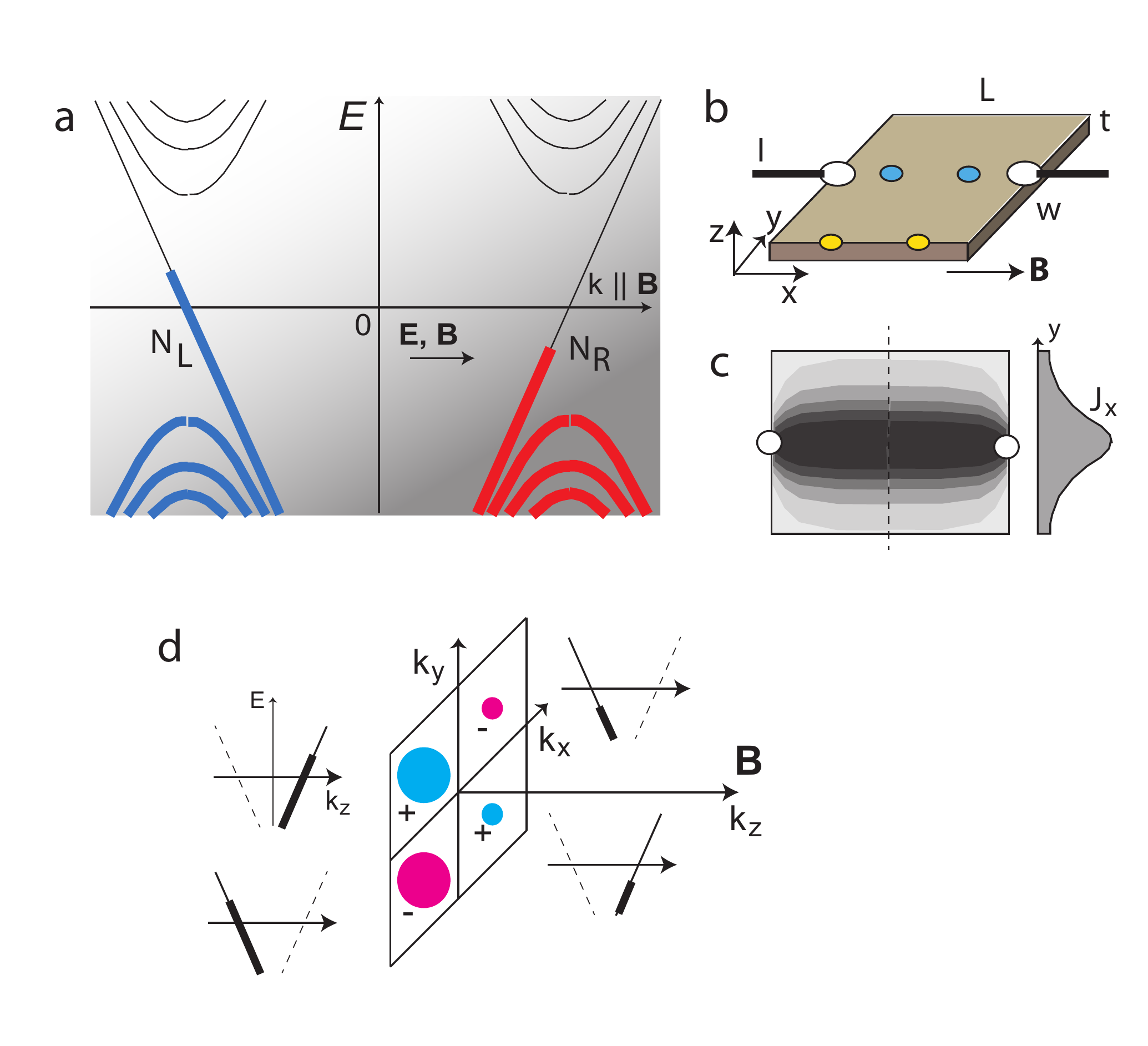}
\caption{\label{figLandau} 
Panel (a): Sketch of the Landau levels of Weyl nodes. In the node with chirality $\chi$ = +1, the chiral level LL0 disperses to the right (velocity $\bf v \parallel B$) while the node with $\chi = -1$ disperses to the left. An electric field $\bf E$ ``pumps'' electrons to the left at a rate given by the anomaly ${\cal A}$. Occupancy of the levels is represented by thick curves. Panel (b) shows the 6 contacts on a plate-like crystal mounted for the squeeze test. The local $E$-fields measured along the spine (blue dots) and along the edge (yellow) define $R_{spine}$ and $R_{edge}$, respectively. Panel (c) shows the heat-map of the current density $J(x,y)$ with $\bf B\parallel E\parallel \hat{x}$. Current jetting concentrates $J(x,y)$ along the spine (dark region) while depleting it at the edges. The profile of $J_x$ vs. $y$ along the dashed line is sketched on the right. Panel (d): Schematic of the chiral zero sound (CZS) at an instant $t$ in the oscillation cycle. The square shows 2 pairs of Weyl nodes in the $k_x$-$k_y$ plane with $\bf B\parallel \hat{z}$ (blue and pink nodes are of chirality $\chi = +1$ and -1, respectively). For each node, the accompanying sketch depicts the chiral dispersion $E$ vs. $k_z$ as a solid line. At instant $t$, the occupancy of the chiral mode (thick lines) is higher in the 2 nodes drawn with larger FS radii. All local charge deviations from equilibrium sum to zero. The CME charge currents also cancel pair-wise. Panels (a), (b) and (c) are from Liang \etal~\cite{Liang}. Panel (d) is based on Ref. \cite{SongDai}.
}
\end{figure}

The CME may have experimental consequences when the system is periodically driven off equilibrium at high frequencies. Song and Dai~\cite{SongDai} have predicted that, in a Weyl semimetal with 2 or more pairs of Weyl nodes, there exists a collective mode in which each Weyl FS executes a breathing mode with a specific phasing between nodes. Unlike a plasmon, this mode displays a gapless, linear dispersion because local charge densities and currents rigorously vanish everywhere at all times. They call the collective mode chiral zero sound (CZS). The simplest example is sketched in Fig. \ref{figLandau}d. The 2 pairs of Weyl nodes are symmetrically arrayed in the $k_x$-$k_y$ plane with ${\bf B}\parallel {\bf k}_z$. As shown, nodes with positive chirality ($\chi = 1$, blue) disperse with velocity ${\bf v} \parallel {\bf B}$, whereas nodes with $\chi = -1$ (pink) disperse with ${\bf v}\parallel -{\bf B}$.

At an instant in the breathing cycle, the local chemical potential $\mu_{loc}$ lies above $E_F$ in the 2 nodes drawn with larger $k_F$ in Fig. \ref{figLandau}d, whereas $\mu_{loc}$ lies below $E_F$ for the nodes with smaller $k_F$. The occupation factor in each node is indicated by thick lines in the accompanying dispersion sketches. With this phasing, it is clear that the charge currents cancel pairwise between the 4 nodes. Deviations of the charge density from equilibrium also cancel. The cancellations allow the mode to propagate as an acoustic wave. The CZS may contribute strongly to both the heat capacity and the thermal conductivity at low $T$ (Sec. \ref{ZS}).

\section{Current jetting artifacts}\label{jetting}
In high-mobility semimetals, longitudinal magnetoresistance experiments are greatly complicated by artifacts caused by current jetting. We assume that the current density  $\langle{\bf J}\rangle$ (spatially averaged over the sample) is $\parallel {\bf B}\parallel \bf\hat{x}$. The cyclotronic motion of the carriers in the $y$-$z$ plane causes the transverse conductivities $\sigma_{yy}$ and $\sigma_{zz}$ to decrease as $1/(\mu_eB)^2$, whereas the longitudinal conductivity $\sigma_{xx}$ is unaffected. The anisotropy results in pronounced concentration of $\bf J$ into a narrow jet aligned with $\bf B$ and a corresponding reduction at the edges parallel to $\bf\hat{x}$. Hence, the voltage drop detected at the edge decreases, masquerading as a negative LMR even though $\sigma_{xx}$ is unchanged.

In the low-mobility semimetals Na$_3$Bi and GdPtBi with $\mu_e\sim$ 1,000 to 2,600 cm$^2$/Vs, the effects of current jetting are still observable, but they introduce a relatively weak distortion that can be corrected for. However, in the Weyl semimetals TaAs, NbAs and NbP, all of which have very high carrier mobilities ($\mu_e>$ 100,000 cm$^2$/Vs), the fractional change in the observed negative LMR signal is typically very small (0.1 to 2 $\%$) and confined to weak $B$ ($\pm$0.5 T). At larger $B$, the MR becomes strongly positive. Frequently, changing the voltage contact placements reverses the sign of the low-$B$ LMR, consistent with an extrinsic origin for the observed LMR.

Using numerical simulations of ${\bf J}({\bf x})$ in a longitudinal $\bf B$, 
Liang \etal~\cite{Liang} devised a ``squeeze'' test that reveals when current jetting artifacts present a serious concern. The test is based on simultaneous measurements of the maximum and minimum values attained by $J_x(x,y,z)$ in the $y$-$z$ plane. The sample is cut in the shape of a thin, square plate with side $a\gg c$, the thickness (Fig. \ref{figLandau}b). Small current injection pads (of diameter $d\simeq c$) are placed in the centers of two opposing edges. The line ($\parallel\bf\hat{x}$) joining the pads is called the spine. Numerical simulations reveal that, under current jetting, $J_x$ is strongly peaked along the spine and suppressed at the edges (Fig. \ref{figLandau}c). Using local voltage contact pairs, the minimum and maximum values of $J_x$ are measured vs. $B$ to yield the effective resistances $R_{edge}(B)$ and $R_{spine}(B)$, respectively. Liang \etal~showed that, in Na$_3$Bi and GdPtBi, $R_{edge}$ and $R_{spine}$ both decrease with increasing $B$, which verifies that the decrease in $\rho_{xx}$ is intrinsic and current jetting effects are subdominant (Secs. \ref{Na3Bi} and \ref{GdPtBi}). 

In contrast, $R_{edge}$ and $R_{spine}$ in high-mobility Bi show divergent field profiles; even at 100 K, $R_{edge}$ decreases to values very close to zero while $R_{spine}$ increases steeply by a factor of 100 (see Fig. 2 in Ref. \cite{Liang}). The divergent trends are artifacts arising from strong focussing of the jet along the spine. The test has also been applied to TaAs (Sec. \ref{TaAs}).

A rule of thumb is to compare $B_Q$ with $B_{cyc}$, the field at which $\mu_eB$ exceeds $\sim 5$. If $B_{cyc}<B_Q$, the steep growth of current-jetting artifacts effectively precludes reliable observation of any intrinsic LMR in the quantum limit.

\section{Chiral anomaly in Magnetoresistance}\label{LMR}
\subsection{The Dirac semimetal Na$_3$Bi}\label{Na3Bi}
The Dirac semimetal Na$_3$Bi crystallizes in the hexagonal $P6_3/mmc$ phase ($D^4_{6h}$). The conduction band is primarily derived from the Na-$3s$ states while the uppermost valence band is derived from Bi-$6p_{x,y}$ states~\cite{WangNa3Bi}. The strong spin-orbit coupling (SOC) causes the Na-$3s$ band to lie below Bi-$6p_{x,y}$ by $\sim$0.7 eV. In addition, SOC lifts the heavy-hole band $|P,\pm\frac32\rangle$ above the light-hole band $|P,\pm\frac12\rangle$. The resulting band crossings lead to Dirac nodes at ${\bf k}^{\pm}_D = (0,0,\pm 0.26 \pi/c)$ along the $z$ axis ($\Gamma$-$A$). Because the $S$ and $P$ bands transform under ${\cal C}_3$ with different irreducible representations and TRI is preserved, the nodes are protected against gap formation. At energies near the node energy $E_0$, it is sufficient to retain the states~\cite{WangNa3Bi}
\be
|S^+, \pm \frac12\rangle, \quad |P^-, \pm \frac32\rangle
\label{SP}
\ee
which are  bonding and antibonding combinations of orbitals centered at the two Na ions (for $S$) and Bi ions (for $P$) with parity eigenvalues $\pm$. 

As described in Sec. \ref{symmetry}, the breaking of TRI in applied $\bf B$ splits each Dirac node into 2 Weyl nodes of opposite chirality $\chi$. Under Landau quantization, the Landau level at $n$ = 0 (LL0) is strictly chiral, dispersing either $\parallel \bf B$ or $-\bf B$ depending on $\chi$. The existence of only 2 Dirac nodes with $E_0$ very close to $E_F$ and the absence of other (spectator) bands at $E_F$ make Na$_3$Bi a very attractive platform to search for the chiral anomaly despite its hyper-sensitivity to moist air.

Notably, in samples with low Na vacancies, $E_F$ lies just below $E_0$. As $T$ decreases from 300 K, the resistivity $\rho$ rises monotonically by a factor of $\sim$20 to saturate at 21 m$\Omega$cm below 20 K, with Hall density $n_H\sim 1\times 10^{17}$ cm$^{-3}$ and mobility $\mu_e\sim$ 2,600 cm$^2$/Vs ~\cite{Xiong}. The distinctly non-metallic profile is also observed in GdPtBi. The weak SdH oscillations observed vs. $B$ indicate that the LL0 is entered at $\sim$6 T. 

\begin{figure}
\centering
\includegraphics[width=15cm]{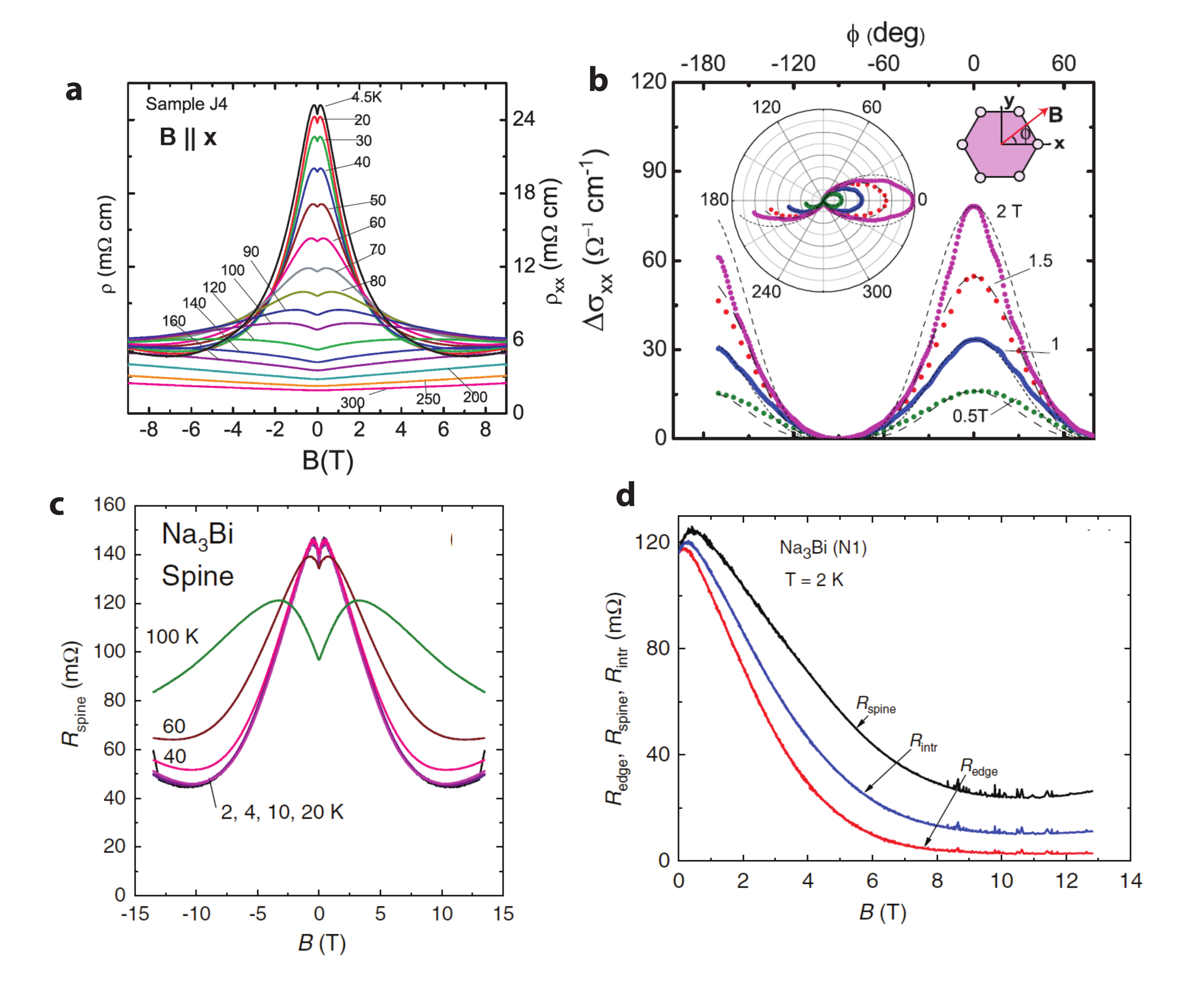}
\caption{\label{figNa3Bi} The chiral anomaly observed in the LMR of Na$_3$Bi. Panel (a) displays magnetoresistance curves measured in Na$_3$Bi in parallel fields (${\bf B}\parallel \langle{\bf J}\rangle\parallel{\bf \hat{x}}$) at selected $T$ from 4.5 to 300 K. Below $\sim$100 K, the steep decrease of the longitudinal resistivity $\rho_{xx}$ in increasing $B$ is direct evidence for the chiral anomaly. Panel (b) shows the variation of the inferred change in conductivity $\Delta\sigma_{xx}$ versus $\phi$ in an in-plane $\bf B$ making an angle $\phi$ with $\bf \hat{x}$, with magnitude $B$ fixed at selected values 0.5 to 2 T. In the polar plot (inset) displaying $\Delta\sigma_{xx}$ (radius) vs. $\phi$, the conductivity enhancement appears as a plume directed along $\bf B$. Application of the squeeze test to distinguish chiral anomaly LMR from current jetting effects.
Panel (c) displays the field profiles of $R_{spine}(B)$ measured in Na$_3$Bi with ${\bf B}\parallel \langle{\bf J}\rangle$ in the squeeze test. If current jetting effects dominate, concentration of $J$ along the spine should lead to an increase in $R_{spine}$ with $B$ (as observed in pure Bi). Here, $R_{spine}$ is observed to decrease instead. Panel (d): The measured profiles of $R_{spine}(B)$ (black curve) and $R_{edge}(B)$ (red) allow the intrinsic curve $R_{int}$ vs. $B$ (blue) to be extracted numerically. In the quantum limit ($B>$ 6 T), $R_{int}$ is $10\times$ smaller than its value in zero $B$. Panels (a) and (b) are adapted from Xiong \etal~\cite{Xiong}; (c) and (d) are from Liang \etal~\cite{Liang}.}
\end{figure}

When $\bf B$ is aligned with $\langle{\bf J}\rangle\parallel {\bf \hat{x}}$, negative LMR becomes apparent at $\sim$100 K (Fig. \ref{figNa3Bi}a). At 4.5 K, $\rho_{xx}$ undergoes a 6-fold decrease before saturating above 8 T, consistent with the appearance of the chiral anomaly (Xiong \etal~\cite{Xiong}). The low mobility (2,600 cm$^2$/Vs) implies that current jetting artifacts should only appear above 11 T~\cite{Liang}. Comparison of the observed $\sigma_{xx}\sim B^2$ in weak $B$ with the Son-Spivak expression yields an estimate of the axial relaxation time $\tau_a\sim$ 40-60 $\tau_0$, the transport lifetime at $B$ = 0.

By tilting $\bf B$ in the $x$-$z$ plane (at an angle $\theta$ to $\bf E$) as well as in the azimuthal $x$-$y$ plane (angle $\phi$), Xiong \etal observed that the conductance enhancement $\Delta\sigma_{xx}(B,\theta,\phi)$ assumes the form of a collimated ``plume'' with axis parallel to $\langle{\bf J}\rangle$ (Fig. \ref{figNa3Bi}b). Displayed as a polar plot, the angular width at 2 T is closer to the form $\Delta\sigma_{xx}\sim \cos^4\theta$ (or $\cos^4\phi$) instead of $\cos^2\theta$. 

Applying the squeeze test (Sec. \ref{jetting}) to Na$_3$Bi, Liang \etal~observed~\cite{Liang} that both resistances $R_{spine}$ and $R_{edge}$ decrease monotonically with increasing $B$ (applied $\parallel\langle{\bf J}\rangle$), confirming that the negative LMR is intrinsic. However, the curve of $R_{spine}(B)$ consistently lies above $R_{edge}(B)$. Moreover, $R_{spine}(B)$ displays a broad minimum near 10 T followed by a gradual increase at larger $B$ as shown in (Fig. \ref{figNa3Bi}c) (this reflects the competition between the intrinsic LMR and current jetting effects which are seen above 10 T). Both features are striking evidence that, despite the low mobility, current jetting effects can still produce observable distortions. Comparing the measured curves against numerical simulations, Liang \etal~showed that it is possible to remove the distortions to extract the intrinsic curve $R_{int}(B)$, which is sandwiched between $R_{spine}(B)$ and $R_{edge}(B)$ (Fig. \ref{figNa3Bi}d). The inferred $R_{int}(B)$ reveals that the chiral anomaly leads to a 10-fold decrease in $\rho_{xx}$, which implies that the axial lifetime $\tau_A$ is 10$\times$ longer than the Drude lifetime $\tau_0$ at $B = 0$. This is currently the most reliable measurement of the ratio $\tau_A/\tau_0$ by dc transport. As seen in Fig. \ref{figNa3Bi}d, $R_{int}$ settles down to a $B$-independent value in the LL0 ($B>$8 T).

This convolution suggests that current jetting distortions may be responsible for the unexpectedly narrow angular width of $\Delta\sigma_{xx}(\theta,\phi)$. Extension to the more elaborate angular MR experiments in tilted $\bf B$ has not been attempted.

\subsection{The Half-Heusler GdPtBi}\label{GdPtBi}
The unit cell of the half-Heusler GdPtBi is comprised of Pt-Gd tetrahedra arrayed in the zincblende structure. The low-lying states involve only the Bi 6$p$ and Pt 4$s$ bands
$|j, m_j \rangle = |\frac32, \pm\frac12\rangle$ and $|\frac32, \pm\frac32\rangle$, 
which are 4-fold degenerate at energy $E_0$ at the $\Gamma$ point. In zero $H$, the lattice symmetry $T_d$ together with TRI protects the 4-fold degeneracy~\cite{Hirschberger,Cano}. In addition to being air-stable, GdPtBi has the advantage that a slight off-stoichiometry during growth (or doping with Au) can shift $E_F$ from below $E_0$ to above.

A magnetic field $\bf B$ lifts the 4-fold degeneracy via the Zeeman energy~\cite{Cano}. The larger Zeeman gap in $|\frac32, \pm\frac32\rangle$ (3$\times$ that in $|\frac32, \pm\frac12\rangle$) leads to band crossings that define Weyl nodes separated in $\bf k$ space. Hirschberger ~\etal~\cite{Hirschberger} observed a large negative LMR when $\bf H$ is aligned with $\bf \langle J\rangle$.

\begin{figure}
\includegraphics[width=15cm]{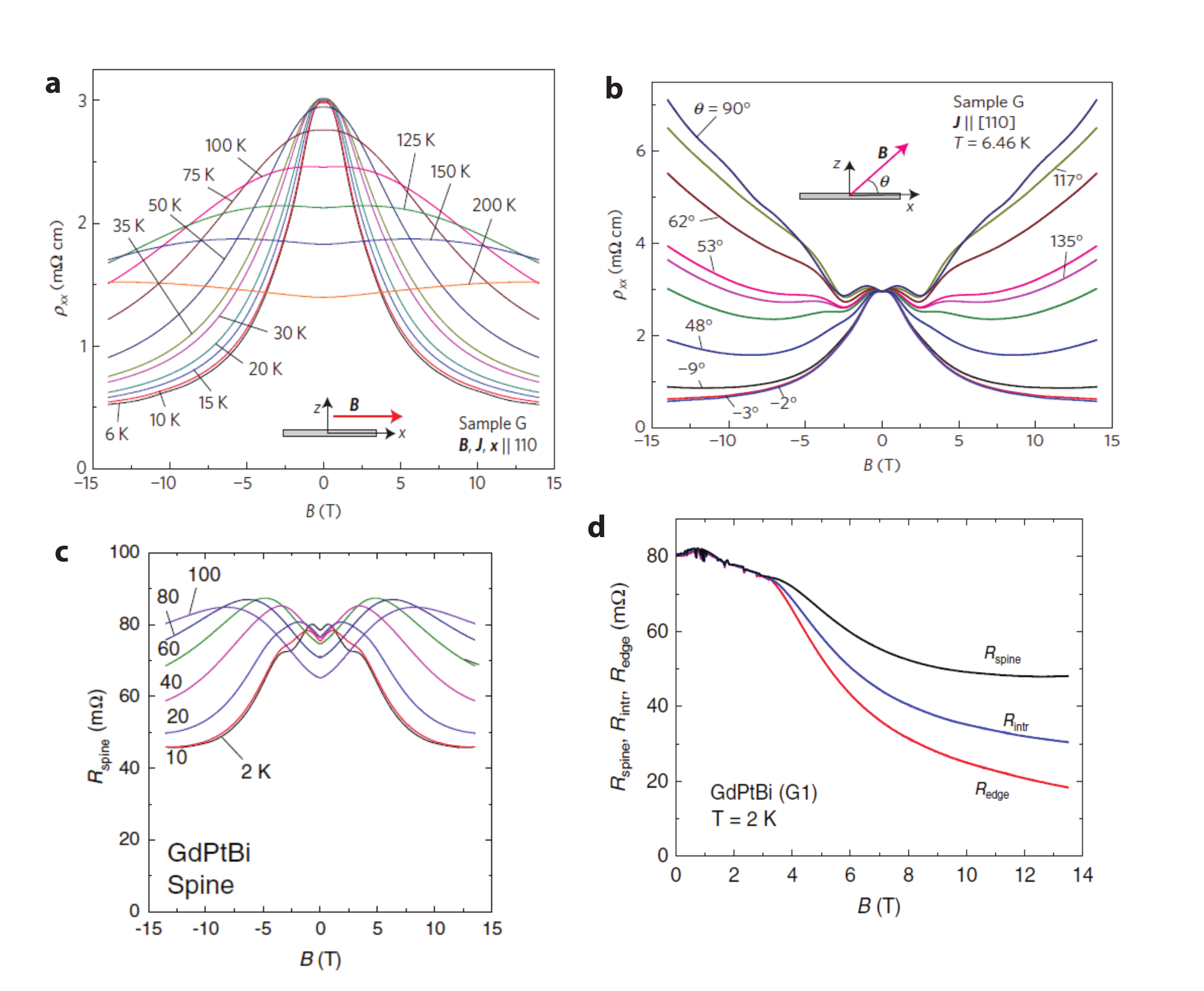}
\caption{\label{figGdPtBi} 
The chiral anomaly in the half-Heusler semimetal GdPtBi. Panel (a) displays the field profiles $\rho_{xx}(B)$ measured with $\bf B\parallel \langle{\bf J}\rangle\parallel hat{x}\parallel (110)$ with $T$ fixed at the values indicated. Below 150 K, a large negative contribution to $\rho_{xx}$ appears. At 6 K, $\rho_{xx}$ displays a bell-shaped profile with a strongly negative LMR closely similar to the profile seen in Na$_3$Bi. Panel (b) shows the effect of tilting $\bf B$ out of the $x$-$y$ plane by angle $\theta$. The negative LMR remains prominent until $\theta$ exceeds $48^\circ$. Panels (a) and (b) are from Hirschberger \etal~\cite{Hirschberger}. Panel (c): Application of the squeeze test to GdPtBi. As in Na$_3$Bi, the resistance on the spine, $R_{spine}(B)$, decreases with increasing $B$, consistent with current jetting effects being subdominant. In combination, the profiles of $R_{spine}$ (black curve) and $R_{edge}$ (red) may be used to obtain numerically the intrinsic resistance $R_{int}(B)$ (blue). In GdPtBi, the quantum limit is reached above 24 T. Panels (c) and (d) are from Liang \etal~\cite{Liang}.
}
\end{figure}

The curves of $\rho_{xx}$ vs. $B$ at fixed $T$ measured with $\bf B\parallel \langle J\rangle$ (Fig. \ref{figGdPtBi}a) bear a close resemblance to those in Na$_3$Bi. As $T$ is lowered from 200 K, the LMR begins to display a prominent negative trend at 125 K. The curve at 6 K displays a steep decrease of $\rho_{xx}$ by a factor of 5 between 0 and 14 T. The angular dependence of the curves of $\rho_{xx}(B)$ was mapped out in detail for $\bf B$ tilted at angle $\theta$ out of the $x$-$y$ plane (Fig. \ref{figGdPtBi}b) and angled within the $x$-$y$ plane. Again, the enhanced conductivity appears as a broad plume centered around the axis with $\bf B\parallel \langle J\rangle$.  

The tunability of $E_F$ by doping provides an important test of the chiral anomaly that could not be done in Na$_3$Bi. By investigating 16 samples with $E_F$ on either side of the Weyl node energy (in zero $H$), Hirschberger ~\etal~\cite{Hirschberger} demonstrated that the LMR is strikingly large ($\rho(9T)/\rho(0) = 0.1$) for samples with $E_F$ close to the Weyl node energy. As $E_F$ is moved away from the node energy (as determined by the weak-field Hall effect), the LMR signal is gradually suppressed. These trends are consistent with the chiral anomaly. 

Liang \etal~\cite{Liang} applied the squeeze test to GdPtBi and observed that both $R_{spine}$ and $R_{edge}$ decreased monotonically with increasing $B$ ($R_{spin}(B)$ is shown in Fig. \ref{figGdPtBi}c). The test confirmed that the negative LMR is intrinsic, just as in Na$_3$Bi. As shown in Fig. \ref{figGdPtBi}d, the numerically extracted intrinsic resistance $R_{int}(B)$ continues to decrease at the highest applied $B$ (the LL0 is reached at $\sim$25 T). To date, Na$_3$Bi and GdPtBi provide the firmest evidence (based on LMR) for the chiral anomaly in semimetals.

\subsection{The layered semimetal ZrTe$_{5}$}\label{ZrTe5}
The semimetal ZrTe$_5$ crystallizes in a layered structure with space group $Cmcm\; (D^{17}_{2h})$. Within each $a$-$c$ layer, prismatic chains of ZrTe$_3$ run parallel to the $a$ axis (the needle axis), with adjacent chains linked by additional Te ions. The quasi-2D layers are stacked along the $b$ axis by van der Waals interaction to form the 3D structure. At $\Gamma$, band inversion of orbitals from Te $p$ states leads to a massive Dirac node above a very small gap.

\begin{figure}
\centering
\includegraphics[width=15cm]{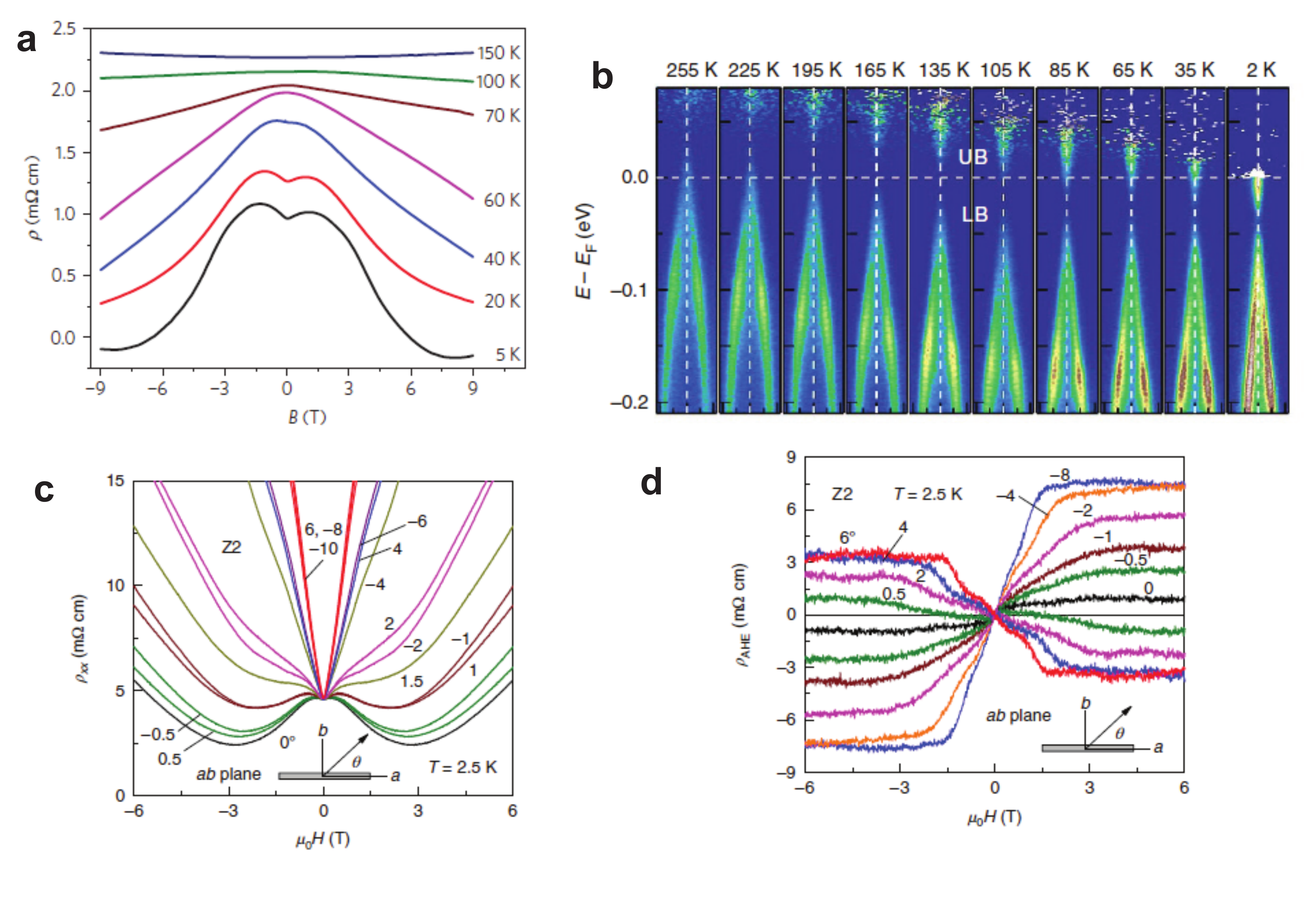}
\caption{\label{figZrTe5} 
Unusual electronic properties of ZrTe$_5$. Panel (a) shows the initial observation of negative LMR  measured at selected $T$ from 5 to 150 K (Li \etal~\cite{QLi}). Panel (b) displays the temperature dependent evolution of the band structure from 2 to 255 K, measured by angle-resolved photoemission along the direction $\Gamma$-$X$ (Zhang \etal~\cite{Zhang2017}). Panel (c) displays the extreme sensitivity of the negative MR to slight tilting of $\bf B$ out of the layer. The negative component of the MR vanishes for tilt angles $|\theta|>1^\circ$ (Liang \etal~\cite{LiangZr}). In addition to the negative MR, ZrTe5 displays a striking array of Berry curvature effects observable in its Hall resistivity $\rho_{yx}$. Panel (d) shows the anomalous Hall effect (AHE) in a tilted magnetic field $\bf B$. Curves of the AHE contribution to the Hall resistivity $\rho_{AHE}$ are plotted vs. $B$ at selected tilt angles $\theta$ between $\bf B$ and the $a$-axis (Liang \etal~\cite{LiangZr}).
}
\end{figure}

Roughly concurrent with the LMR experiment on Na$_3$Bi~\cite{Xiong}, negative LMR was also reported in ZrTe$_{5}$ by Li~\etal~\cite{QLi}, and attributed to the chiral magnetic effect (Fig. \ref{figZrTe5}a). Subsequent experiments on ZrTe$_5$ have led to a bewildering array of transport behavior, caused by the sensitivity of the unusually small carrier population to Te vacanies and variation of the chemical potential $\mu$ with $T$. In early samples grown by chemical vapor transport with high densities of Te vacancies, the resistivity profile $\rho_a$ vs. $T$ displays a large peak at $T_p$ (varying from 60 to 160 K). Angle-resoved photoemission measurements (Fig. \ref{figZrTe5}b) on a sample with $T_p$ = 135 K show that $\mu$ shifts considerably with $T$~\cite{Lifshitz}. At 2 K, $\mu$ lies at the bottom of the conduction band close to $\Gamma$. As $T$ is raised to 255 K, $\mu$ crosses the small gap (50-80 meV) to end up near the top of its valence band (at $T_p$, $\mu$ is mid-gap, in agreement with the profile of $\rho_a$). Evidence for a temperature-driven topological transition from a strong TI to weak TI occuring at $T_p$ (138 K) has been obtained from infrared spectroscopy~\cite{Bernhard}.

The Te vacancy density is lower in flux-grown crystals. As $T$ decreases from 300 to 2 K, $\rho_a$ rises monotonically to approach saturation below $\sim$10 K. ARPES measurements~\cite{LiangZr} show that, at 17 K, $\mu$ lies close to the top of the valence band ($\mu$ does not enter the band gap in the sample studied). The existence of a moderately large ($\sim 50\%$) negative LMR was confirmed in measurements with $\bf B\parallel \langle J\rangle\parallel a$~\cite{LiangZr}. However, when $\bf B$ was tilted at an angle $\theta$ exceeding 1.5$^\circ$, the negative LMR vanished as shown in Fig. \ref{figZrTe5}c (with $\theta$ defined in the inset). In addition, the LMR monitored at $\theta = 0$ changed to a positive sign when $B$ was increased above 3 T. Both the high sensitivity to $\theta$ and the sign change are not understood. Unlike Na$_3$Bi and GdPtBi, the squeeze test to eliminate current jetting artifacts has yet to be performed. The high mobility (60,000 cm$^2$/Vs at 2 K), which makes $B_{cyc}\ll B_Q$ (0.83 vs. 4 T), suggests that current jetting effects may be dominant in ZrTe$_5$.

Apart from the LMR behavior, ZrTe$_5$ displays a rich assortment of transport features at 2 K engendered by the Berry curvature ${\boldsymbol\Omega}$. As shown in Fig. \ref{figZrTe5}d, the Hall resistivity $\rho_{yx}$ displays a large anomalous Hall effect (AHE) which has been mapped out over the entire solid angle of the vector $\bf H$ by Liang \etal~\cite{LiangZr}. A very interesting feature is the emergence of a true, anomalous planar Hall effect that reverses sign with the in-plane $\bf H$.  The sign reversal with $\bf H$ (Onsager behavior) distinguishes it from the so-called ``planar Hall effect'' engendered by domain wall anisotropy in magnetic thin films.

Under uniaxial stress applied $\parallel\bf a$, $\rho_a$ in flux-grown crystals initially decreases to a attain a minimum at a strain $\epsilon_{min}\sim\; 0.12\,\%$ and then rises quadratically at higher strain. Mutch~\etal~\cite{Mutch} interpret this unusual behavior as evidence for a sign-change of the mass term $m$ in the Dirac cone arising from a topological phase transition from a strong TI to a weak TI phase. A negative LMR that changes sign above $\sim$2 T and is highly sensitive to $\theta$ was also observed in the two TI phases.

\subsection{Weyl Semimetals}\label{TaAs}
\begin{figure}
\centering
\includegraphics[width=15cm]{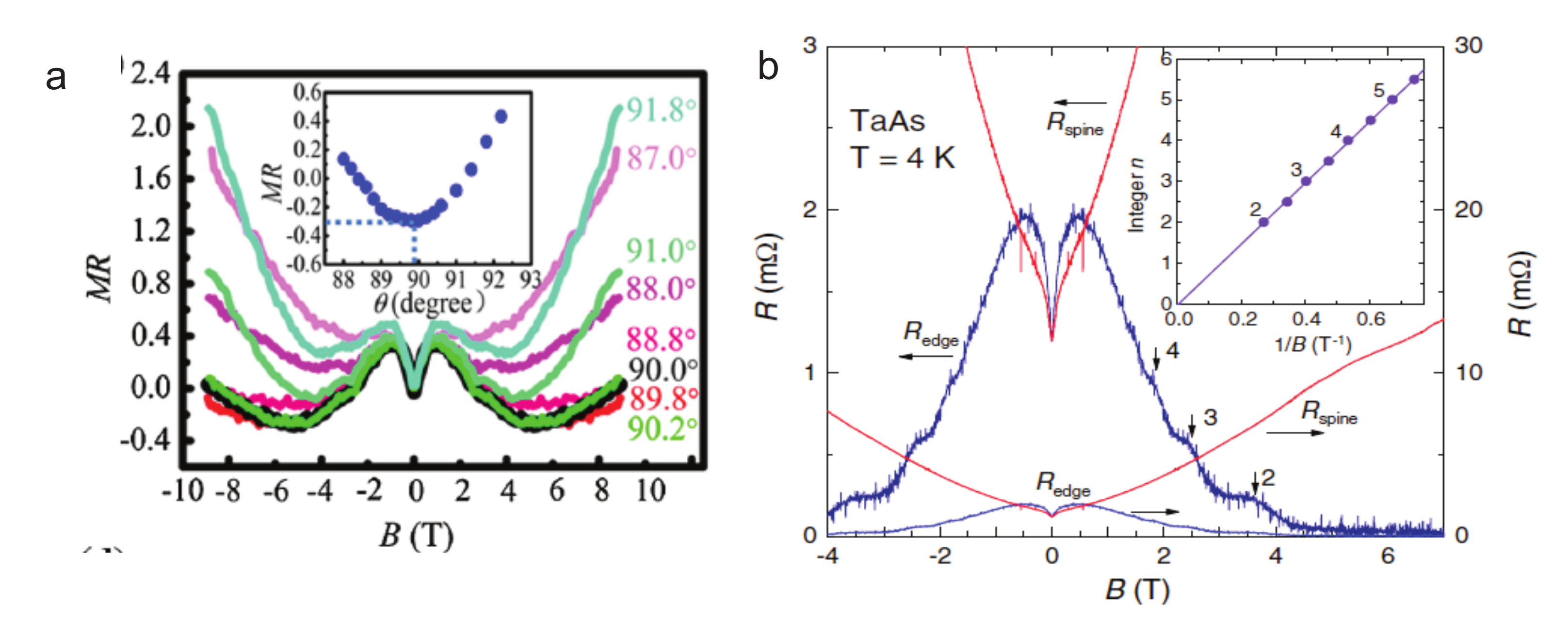}
\caption{\label{figTaAs} Panel (a): The observed fractional change in resistivity, MR $\equiv (\rho(B)-\rho(0))/\rho(0)$, observed in TaAs at 1.8 K~\cite{Huang2015}. The negative MR is highly sensitive to slight tilts of $\bf B$ away from $\bf E$. The negative MR vanishes when $\theta$ deviates from $90^\circ$ by 2$^\circ$. Panel (b) shows the results of applying the squeeze test to TaAs at 4 K~\cite{Liang}. The edge and spine resistances, $R_{edge}$ and $R_{spine}$, respectively, strongly diverge once $B$ deviates from zero (as shown by the arrows, both curves are displayed on two scales). This implies that the observed LMR is dominated by current jetting artifacts because of the very high carrier mobility. The inset shows the index plot of LLs derived from the weak oscillations in $R_{edge}$ which yields $B_Q$ = 7.04 T. 
}
\end{figure}

Unlike in the Dirac semimetals, the space group of the Weyl semimetals, TaAs, NbAs, TaP and NbP, lacks inversion symmetry. Hence each Dirac node is already split into isolated Weyl nodes in zero $\bf B$. TaAs belongs to the space group $I4_1md$. There exist 24 Weyl nodes, with 8 located on the $k_z = 0$ plane and 16 lifted off the plane. The large number of nodes combined with the very high mobilities (100,000 to 150,000 cm$^2$/Vs) make LMR data difficult to analyze. Initially, observations of a small, negative LMR in weak $H$ were interpreted as the chiral anomaly (Fig. \ref{figTaAs}a)~\cite{Huang2015,Shuang}. However, subsequent studies~\cite{Hassinger,ZhuanXu} found that the LMR feature is fragile, changing sign if the voltage contacts are rearranged.

The application of the squeeze test to TaAs and NbP demonstrated~\cite{Liang} that current jetting distortions onset at a field $B_{cyc}\sim$0.5 T far below $B_Q$ = 7.04 T ($B_Q$ is determined by the quantum oscillations). Figure \ref{figTaAs}b shows that, as $B$ is increased from zero, $R_{spine}$ increases very steepy whereas $R_{edge}$ falls to values below the detection limit before $B_Q$ is attained (the divergent profiles are closely similar to those in Bi). Based on these findings, Liang \etal~conclude that LMR experiments cannot be used to confirm the chiral anomaly in TaAs and NbP. Alternate techniques to detect the anomaly-induced currents are described below.

\section{Complementary Experiments}\label{related}

\subsection{Non-local Transport}
\begin{figure}
\centering
\includegraphics[width=15cm]{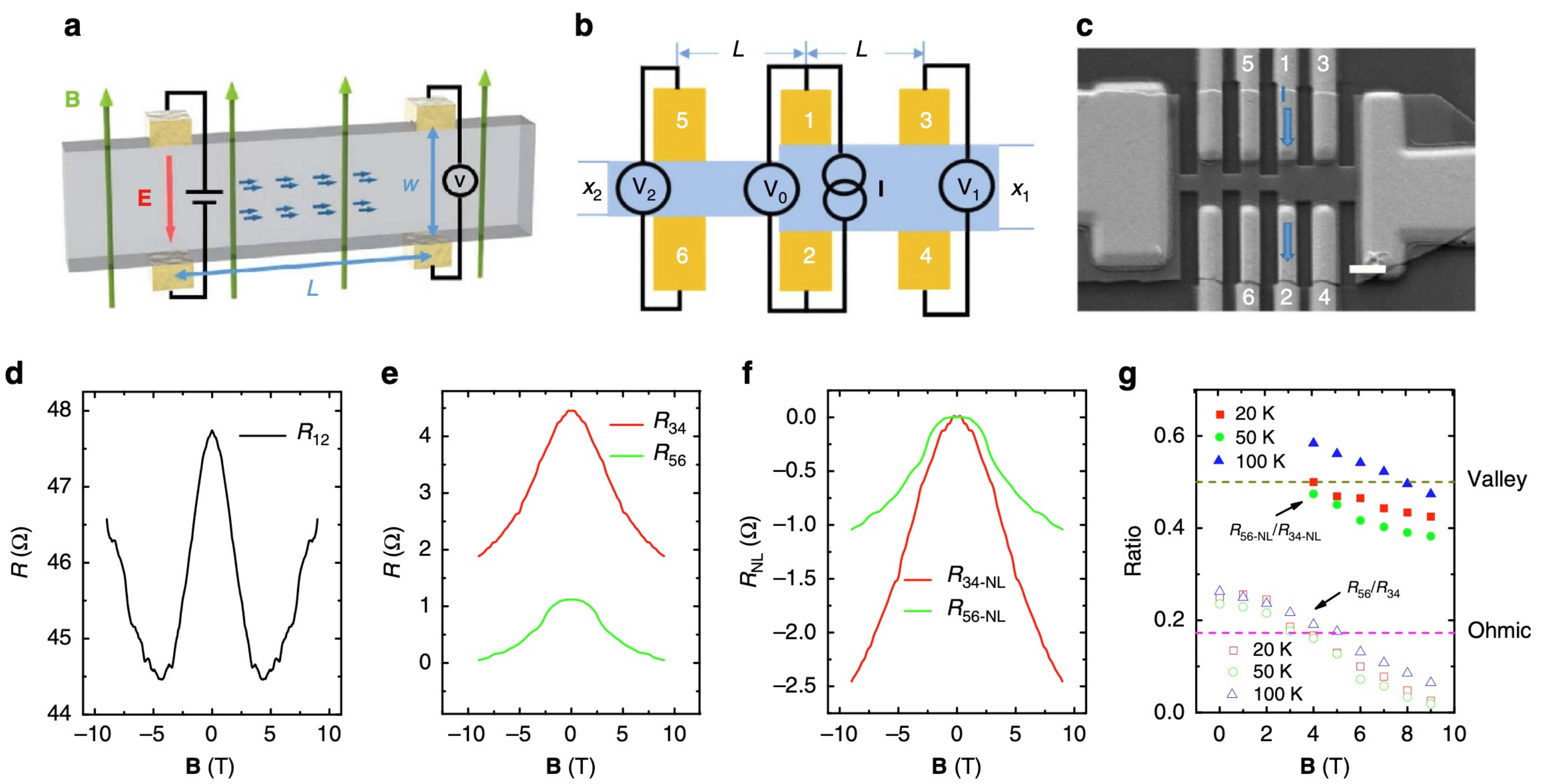}
\caption{\label{fignonlocal} Detection of non-local signal in the Dirac semimetal Cd$_3$As$_2$. Panel (a): Schematic view of the valley diffusion process. Parallel (antiparallel) $\bf E$ and $\bf B$ fields generate the charge imbalance between two Weyl nodes due to the chiral anomaly. The charge imbalance of different valleys can diffuse across the sample and be converted into a nonlocal voltage along the direction of $\bf B$. Panel (b): Schematic view of the nonlocal resistance measurement with different diffusion channel width. Current is applied through terminal 1–2, while terminals 3–4 and 5–6 are used to measure the nonlocal resistance. The diffusion length $L$ is 2 mm. Panel (c): The electron micrograph of the device (white scale bar is 2 mm). The contact regime in terminals 3–4 is slightly larger than that of 5–6. Panel (d): The two- terminal local resistance (R$_{12}$) at 20 K. Panel (e): The nonlocal resistance (R$_{34}$ and R$_{56}$) at 20 K. Panel (f): The pure nonlocal resistance (R$_{34-NL}$ and R$_{56-NL}$) after subtracting the Ohmic diffusion at 20 K. Panel (g): Resistance ratio of R$_{34}$/R$_{56}$ and R$_{34-NL}$/R$_{56-NL}$ versus $B$ at different $T$. Dashed lines labelled as Valley and Ohmic correspond to ratios 0.50 and 0.17, respectively. From Zhang \etal~\cite{Zhang2017}.}
\end{figure}

Parameswaran \etal~\cite{Sid} have proposed a non-local transport experiment to detect the chiral anomaly in Dirac/Weyl semimetals. The test assumes that an unbalance in the electrochemical potentials $\mu^{R,L}_{EC}$ of two Weyl nodes (labelled $R$ and $L$) is established by injection of a pump current at one end ($x = 0$) of a long thin-film sample in an applied probe field ${\bf B}_p$. The electro-chemical potential difference, $\delta\mu_{EC}(x) = \mu^R_{EC}(x) - \mu^L_{EC}(x)$, decays exponentially along the sample's length as $\delta\mu_{EC}(x) = \delta\mu_{EC}(0)e^{-x/\ell_v}$. The diffusion length $\ell_v$ is given by $\ell_v = \sqrt{D\tau_v}$, where $D$ is the carrier diffusion constant and $\tau_v$ is the intervalley scattering lifetime (Fig. \ref{fignonlocal}a). 
The electro-chemical potential unbalance may be detected by pairs of voltage contacts at various points along the sample (Fig. \ref{fignonlocal}b). Importantly, the chiral current direction of either Weyl node depends on the local direction of the magnetic field. Hence, if one could apply a detection field ${\bf B}_d$ at the voltage contacts distinct from ${\bf B}_p$, one could verify if the detected signal has a sign dictated by ${\bf B}_d\cdot{\bf B}_p$.

In Ref. \cite{Zhang2017}, Zhang \etal~employed a focused ion beam to fabricate multiple pairs of probes on a CVD grown thin film Cd$_3$As$_2$ sample to measure the non-local signal (Fig. \ref{fignonlocal}c). They showed that the measured non-local signal has contributions from both Ohmic diffusion (the conventional current) and the polarization diffusion arising from the charge pumping (Figs. \ref{fignonlocal}d,e,f). They found that the valley polarization diffusion length is $\sim 3\times$ the conventional Ohmic diffusion length (Fig. \ref{fignonlocal}g). Because of the short length scales, however, the crucial test of applying an independent ${\bf B}_d$ at the voltage contacts could not be carried out.

\subsection{Thermopower}\label{thermopower}
The thermopower $S_{xx}$ provides a probe of the LL0. Instead of the usual quadratic dispersion along $\bf H\parallel\hat{x}$, we now have a strictly 1D linear dispersion. By the Mott formula, the flat density of states causes strong suppression of $S_{xx}$.
As noted, the lowest Landau level (LL0) of Weyl fermions is chiral. This striking feature implies that the density of states ${\cal D}_0(E)$ is nominally independent of energy $E$, in contrast with the divergent form ${\cal D}_n\sim (E-E_n)^{-\frac12}$ at higher LLs (with $E_n = (n+\frac12)\hbar\omega_c$). 

The thermopower vanishes if particle-hole symmetry exists at the chemical potential. Using the Mott relation, 
$S_{xx}=(\pi^2 k_B^2 T/3e)[\partial \ln \sigma_{xx}/\partial E]$, 
we have $S_{xx}\sim \partial {\cal D}_n/\partial E$ 
if the $E$ dependence of the carrier lifetime is negligible. Hence $S_{xx}$ is expected to be strongly suppressed in the chiral LL0.

\begin{figure}
\centering
\includegraphics[width=15cm]{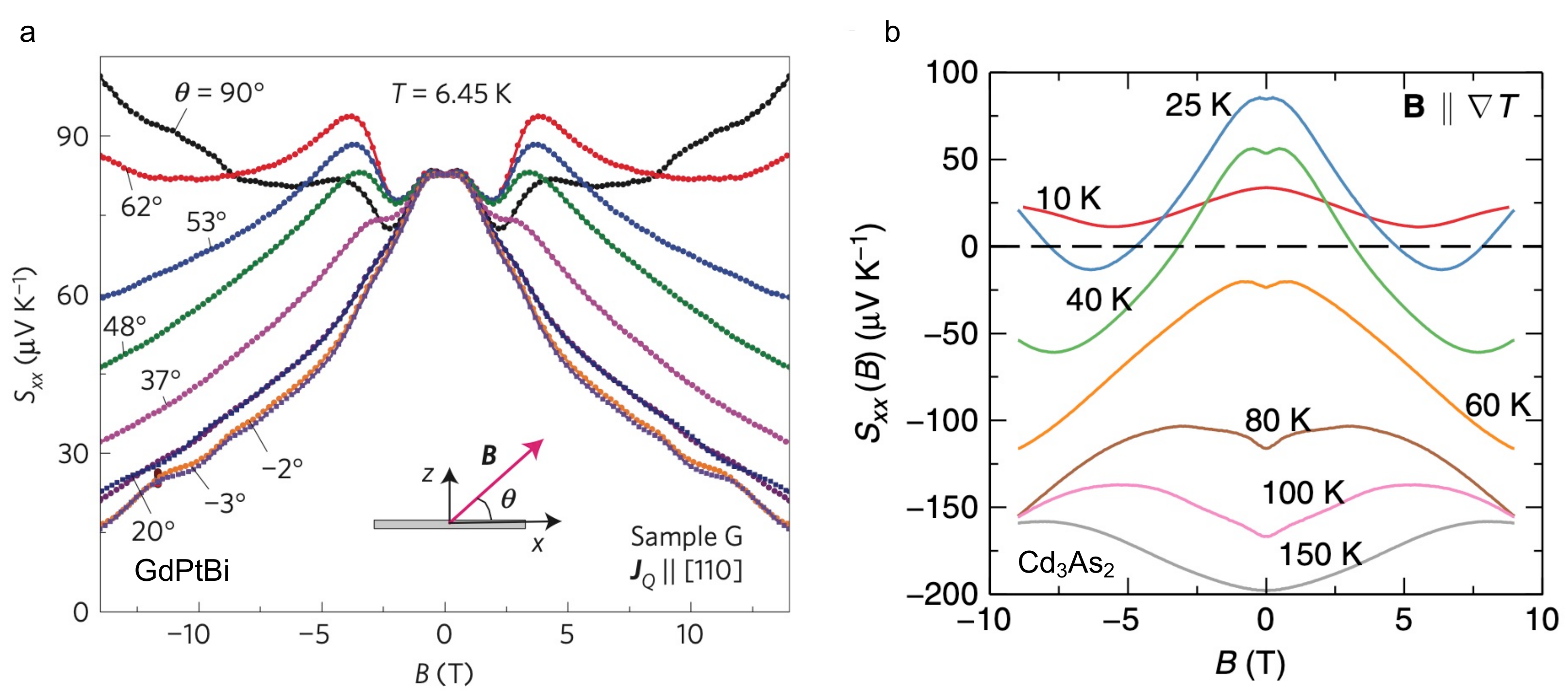}
\caption{\label{figthermo}
Curves of the thermopower $S_{xx}$ vs. $B$ observed in GdPtBi (Panel a) and in Cd$_3$As$_2$ (Panel b). Panel (a) shows curves of $S_{xx}(B)$ measured in GdPtBi at selected tilt angles $\theta$ of $\bf B$ relative to the $x$-$y$ plane at $T$ = 6.45 K. The thermal current density $J_Q$ is applied along the [110] direction. The pronounce decrease of $S_{xx}$ with increasing $B$ at $\theta = 0$ is consistent with the increased dominance of the lowest (chiral) Landau level in which $S_{xx}$ is strongly suppressed (see text). Weak SdH oscillations are observed for the curve at $\theta = 0$. Panel (b) plots $S_{xx}(B)$ in Cd$_3$As$_2$ measured at selected values of $T$. As the temperature decreases below 100 K, the curves become increasingly negative. Panel (a) is adapted from Hirschberger \etal~\cite{Hirschberger}. Panel (b) is from Jia \etal~\cite{Jia2016}.}
\end{figure}

Hirschberger \etal~\cite{Hirschberger} investigated the strong variation of the thermopower $S_{xx}(B,T)$ in GdPtBi vs $B$. When ${\bf B}$ is aligned $\parallel \langle{\bf J}_Q\rangle$ (the applied thermal current density), $S_{xx}$ is observed to decrease monotonically by a factor of 6 as $B$ is increased from 0 to 14 T in a sample with $B_Q\sim$25 T (Fig. \ref{figthermo}a). The steep decrease is consistent with the approach to the strongly suppressed value as $\mu$ approaches the quantum limit.

Unlike in Na$_3$Bi, a complication in GdPtBi is that the Weyl nodes have to be created by applying a finite $\bf B$~\cite{Hirschberger,Cano}. The Weyl states and their characteristic Landau levels appear only when $B$ exceeds $B_p\sim$ 4 T. The field-creation process becomes manifest if we compute the ``thermoelectric conductivity'' $\alpha_{xx} = S_{xx}/\rho_{xx}$, which relates the charge current to the applied gradient via $J_x = \alpha_{xx}(-\nabla T)$. The field profile of $\alpha_{xx}$ displays a prominent peak at $B_p$, which marks the onset of the Weyl regime ($B>B_p$). Regardless of this distinction, $\alpha_{xx}$ is also observed to decay in magnitude when $B$ exceeds $B_p$.
 
The absence of Weyl fermions in GdPtBi in the low-field region $0<B<B_p$ precludes comparison of the measured curves with weak-field Boltzmann-equation calculations of $\alpha_{xx}(B)$ reported in Refs. \cite{Lundgren,Sharma,SpivakAndreev}. The suppression of $S_{xx}$ by $B$ has also been observed in Cd$_3$As$_2$ by Jia \etal ~\cite{Jia2016}. Curves of $S_{xx}$ vs. $B$ (with ${\bf B}\parallel \langle{\bf J}_Q\rangle$) are shown in Fig. \ref{figthermo}b.

\subsection{Thermal Conductivity and chiral zero sound}\label{ZS}
\begin{figure}
\centering
\includegraphics[width=15cm]{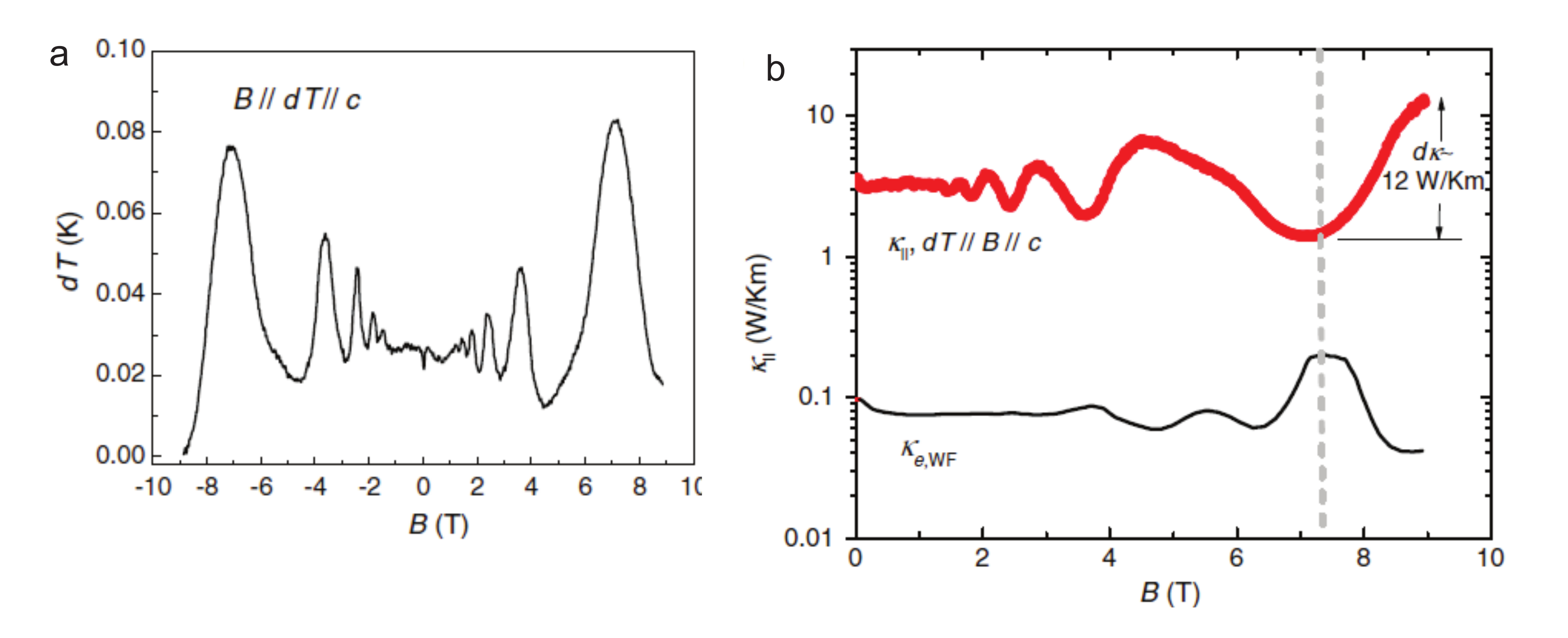}
\caption{\label{figzerosound} Panel (a): Raw experimental trace of the temperature gradient $dT(B)$  in the semimetal TaAs as the field $B$ is swept from -9 to 9 T at constant heater power with ${\bf B}\parallel \langle{\bf J}_Q\rangle\parallel {\bf c}$ and $T$ fixed at 2 K. Giant quantum oscillations are observed in the longitudinal thermal conductivity $\kappa_{\parallel}$. Panel (b): The amplitudes of the quantum oscillations in $\kappa_{\parallel}$ at low $T$ (red curve) are nearly two orders of magnitude larger than the oscillations in the thermal conductivity $\kappa_{e,WF}$ inferred from the longitudinal conductivity $\sigma_{\parallel}$ using the Wiedemann-Franz law (black curve). From Xiang \etal~\cite{Xiang2019}.}
\end{figure}

In a recent measurement of the thermal conductivity $\kappa_{xx}$ vs. $B$ in TaAs, Xiang \textit{et al.} \cite{Xiang2019} observed magnetic quantum oscillations in $\kappa_{xx}$ with remarkably large amplitude when ${\bf J}_Q$ is aligned with $\bf B$ (Figs. \ref{figzerosound}a). The oscillations have the same period as (but are antiphased with) the SdH oscillations in the conductivity $\sigma_{xx}$. The peak-to-peak amplitude $\Delta\kappa_{xx}\sim$ 12 W/Km of the largest oscillation is 3.4$\times$ the zero-$B$ value $\kappa$ (Figs. \ref{figzerosound}b). The overall values of $\kappa_{xx}$ are 50-100$\times$ larger than predicted by the standard Wiedemann-Franz law. After eliminating several potential causes, the authors identify the oscillations as arising from the propagation of the CZS mode predicted by Song and Dai~\cite{SongDai} (Sec. \ref{CME}). They show that both the magnitude and phase of the oscillations can be fitted to the CZS model.

\subsection{Optical pump-probe experiment}\label{optical}
\begin{figure}
\centering
\includegraphics[width=15cm]{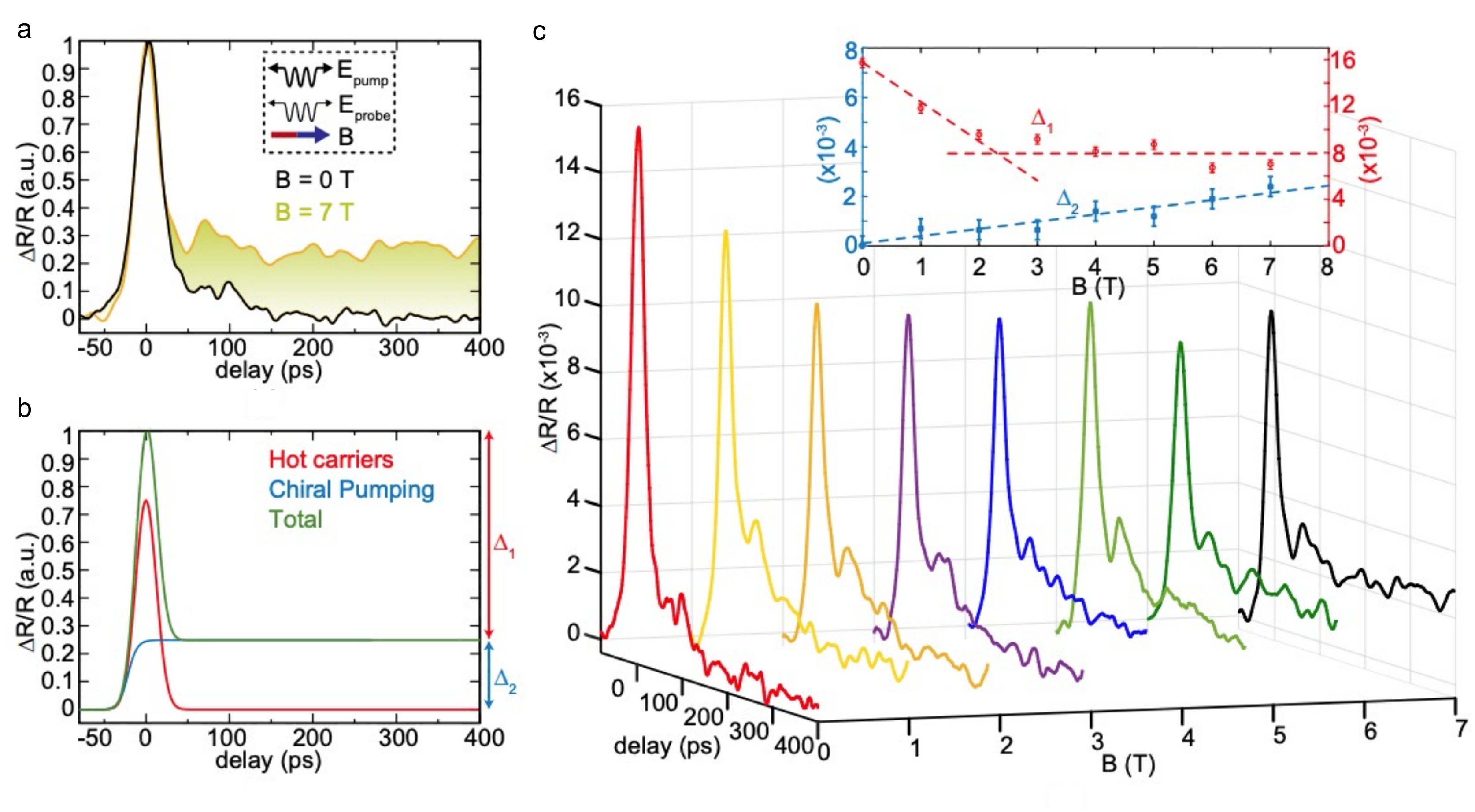}
\caption{\label{figoptical} 
Optical detection of long-lived current induced by charge pumping in the Weyl semimetal TaAs. 
Panel (a) shows the pump-induced fractional increase in probe reflection $\Delta R/R$ at zero magnetic field (black curve) and at $B$ = 7 T (yellow) as a function of the time delay between the pump and probe pulses, with ${\bf E}_{pump}\parallel {\bf E}_{probe}\parallel {\bf B}$. The prominent peak close to zero time delay, common to both traces, tracks the fast relaxation of hot carriers. However, when $B$ is set at 7 T, the pump-probe trace exhibits a component (shaded area) that relaxes on a time scale far longer than 400 ns. The long-lived component is associated with the chiral anomaly. Panel (b): Simulated pump-probe traces for fast hot carriers effects (red), metastable chiral pumping (blue), and the net result (green). $\Delta_1$ is defined as the maximum pump-induced change in probe reflection near zero time delay, and $\Delta_2$ characterizes the long-lived component. Panel (c) shows 8 pump-probe traces as $B$ is set at successively larger values. The inset shows $\Delta_1$ (red) and $\Delta_2$ (blue) vs. $B$.
From Jadidi \etal~\cite{Jadidi2019}.}
\end{figure}

Jadidi \textit{et al.} \cite{Jadidi2019} have employed pump-probe measurements at terahertz frequencies to investigate chiral charge pumping and measure carrier relaxation in the Weyl semimetal TaAs in applied $\bf B$. An intense ``pump'' pulse at frequency 3.4 THz with optical field ${\bf E}_{pump}\parallel \bf B$ is employed to generate photo-excited carriers. The low photon energy (14 meV) ensures that only carriers very close to the Weyl nodes are excited. The resulting changes to the reflection coefficient are detected by a weak probe pulse. By varying the delay time between pump and probe, they infer the relaxation time of the carriers (Fig. \ref{figoptical}). In addition to the usual hot-carrier contribution (which relax very rapidly), they observed a long-lived metastable response that persists beyond 1 ns in the presence of $\bf B$ (Fig. \ref{figoptical}a). The metastable response is interpreted as evidence for the axial current (Fig. \ref{figoptical}b). As a test, they verified that the metastable response is present only when ${\bf E}_{pump}$ is aligned parallel to $\bf B$, and vanishes when ${\bf E}_{pump}$ is perpendicular to $\bf B$. The metastable axial anomaly signal is found to be linear in $B$ (inset in Fig.\ref{figoptical}c).

\section{Perspective}\label{perspective}
In this review, we surveyed experiments on the chiral anomaly in condensed matter physics. Our discussion, however, falls short of conveying the importance of the anomalies in quantum field theory. The experiments here join a very long thread of anomaly topics that extends through meson physics to quark physics and gravitation. In the Supplementary Information, we provide a longer discussion of the crucial discovery of Adler and Bell and Jackiw of the chiral anomaly in the context of the decay of $\pi^0$. We describe how the anomaly ${\cal A}$ wrecks the conservation of the axial current ${\bf J}^5$. At the level of quarks, the (non-Abelian) anomaly solves the $U_A(1)$ problem and leads to the $\theta$-vacuum. At a higher level of abstraction, we refer to Fujikawa's path-integral formulation which relates ${\cal A}$ to the chiral zero-modes of massless fermions. The anomaly has a starring role in the long mathematical odyssey from the Gauss-Bonnet theorem to the Atiyah-Singer index theorem. We direct readers to several references on this rich subject.


\newpage

\setcounter{equation}{0}

\section{Supplementary Information: Chiral Anomaly in QFT}\label{QFT}
We briefly survey the discovery of the chiral anomaly in quantum field theory (QFT). Introductory discussions of anomalies are found in Peskin and Schroeder~\cite{Peskin} and Cheng and Li~\cite{ChengLi}. More advanced treatments are given by Weinberg~\cite{Weinberg}, Nakahara~\cite{Nakahara} and Bertlmann~\cite{Bertlmann}.  Peskin and Schroeder~\cite{Peskin}, Cheng and Li~\cite{ChengLi}, and Aitchison and Hey~\cite{Aitchison} are excellent references for QCD phenomenology. Recent brief reviews of anomalies are given by Adler~\cite{AdlerFifty} and Jackiw~\cite{Jackiw}. Concepts in differential geometry are reviewed in Nakahara~\cite{Nakahara} and Frankel~\cite{Frankel}.

\subsection{Chiral symmetry of massless fermions}

The Lagrangian describing the free Dirac fermion is
\be
{\cal L} = \bar\Psi(i\gamma^\mu\partial_\mu -m)\hat{\Psi},
\label{LD}
\ee
where $m$ is the fermion mass and $\hat{\Psi}$ is a 4-spinor (with $\bar{\Psi} \equiv \hat\Psi^\dagger\gamma^0)$. The set of $4\times 4$ Dirac matrices, $\gamma^\mu$, are $\gamma^0, \gamma^1, \gamma^2, \gamma^3$. Here, we focus on the chirality matrix $\gamma_5=\gamma^5 \equiv i\gamma^0\gamma^1\gamma^2\gamma^3$ which anticommutes with $\gamma^\mu$, viz.
\be
\{\gamma_5,\gamma^\mu\}=0, \quad (\mu = 0,1,2, 3).
\label{anti}
\ee
In the Dirac (or Bjorken-Drell) representation, $\gamma^0$ is diagonal whereas $\gamma_5$ is block off-diagonal.

Weyl observed that, in the limit $m\to 0$, the Lagrangian ${\cal L}$ becomes $({\cal L}_L + {\cal L}_R)$, which describes independent ``left'' and ``right'' massless populations. In this limit, it is more convenient to adopt the chiral representation in which $\gamma_5$ is diagonal, viz. $\gamma_5 =
				\left[\begin{array}{cc}
						-\hat{1} & 0\\
								0  & \hat{1} \end{array}\right].
								$
The left- and right-handed spinors $\hat\Psi_{L,R}$ are the eigen-spinors of $\gamma_5$, with eigenvalues $\chi \equiv \pm 1$, viz.
\be
	\gamma_5\hat{\Psi}_L = (-1)\hat{\Psi}_L,
	\quad \gamma_5\hat{\Psi}_R = (+1)\hat{\Psi}_R.
\label{gam5}
\ee
In the chiral representation, we have
\be
\hat{\Psi}_L = \left(\begin{array}{c}
						\hat{u}_L\\
								0 \end{array}\right),
	\quad\hat{\Psi}_R = \left(\begin{array}{c}
						0\\
								\hat{u}_R
								\end{array}\right),
\label{PsiL}
\ee
where $\hat{u}_{L,R}$ are 2-spinors. The labels $L$ and $R$ warrant a comment. In photons, right- and left-handedness refer to the locking of the photon's spin $\vec{\sigma}$ parallel or anti-parallel to its momentum $\bf p$. This actually refers to the eigenvalues $h$ of the helicity operator $\vec{\sigma}{\hat{\bf\cdot p}}$. For massless fermions, we have $\chi = h$ for fermions. However, $\chi = -h$ for antifermions. Hereafter, `left' and `right' are understood as labels for the eigenstates of $\gamma_5$, rather than the helicity.

It is helpful to regard $\hat\Psi_{L,R}$ as produced by the projection operators ${\cal P}_\pm = \frac12(1\pm \gamma_5)$ satisfying ${\cal P}_\pm^2={\cal P}_\pm$ and ${\cal P}_-{\cal P}_+=0$, viz.
\be
\hat{\Psi}_L = \frac{(1- \gamma_5)}{2}\hat{\Psi},
\quad \hat{\Psi}_R = \frac{(1+\gamma_5)}{2}\hat{\Psi}.
\label{proj}
\ee

Projection yields $(1-\gamma_5)\gamma^0\gamma^\mu(1+\gamma_5) = 0$. This verifies Weyl's observation that, if $m\to0$, ${\cal L}$ is the sum of independent $L$ and $R$ terms, viz.
\be
{\cal L} = i\bar\Psi_L \gamma^\mu\partial_\mu \hat{\Psi}_L +
i\bar\Psi_R \gamma^\mu\partial_\mu \hat{\Psi}_R.
\label{LDLR}
\ee

By Noether's theorem, symmetries of the Lagrangian ${\cal L}$ lead to conservation laws. With $m=0$ in Eq. \ref{LD}, ${\cal L}$ is invariant under the 2 global transformations
\be
\hat{\Psi} = \left(\begin{array}{c}
						\hat{u}_L\\
								\hat{u}_R \end{array}\right)
								\to \left(\begin{array}{c}
						\hat{u}_L\\
								\hat{u}_R \end{array}\right)e^{i\theta}, \quad \quad
\left(\begin{array}{c}
						\hat{u}_L\\
								\hat{u}_R \end{array}\right)
								\to \left(\begin{array}{c}
						\hat{u}_L e^{i\theta}\\
								\hat{u}_R e^{-i\theta} \end{array}\right).
\label{rotate}
\ee
In the first transformation ($\hat{\Psi}\rightarrow e^{i\theta}\hat{\Psi}$), the 2-spinors $\hat{u}_L$ and $\hat{u}_R$ are rotated in isospin space by the same angle $\theta$, while in the second ($\hat{\Psi}\rightarrow e^{i\theta\gamma_5}\hat{\Psi}$), the rotations are of opposite signs (axial rotation).

The two types of rotations are starting points for discussing the symmetry properties of massless fermions. Using the chiral representation, we either rotate the phase of $\hat{u}_L$ and $\hat{u}_R$ in unison, or in opposite directions (the latter is achieved by the chirality matrix $\gamma_5$ in the exponent). If the Lagrangian is invariant under both operations, we have conservation of the vector current $j^\mu$ and the axial (or chiral) current $j^\mu_5$ defined by
\be
j^\mu =  \bar{\Psi}\gamma^\mu \hat{\Psi},
\quad j^\mu_5 = \bar{\Psi}\gamma^\mu \gamma_5\hat{\Psi}.
\label{jj}
\ee
With $\hat{\Psi} = \hat{\Psi}_L+\hat{\Psi}_R$, $j^\mu$ simplifies to the sum of vector currents associated with $L$ and $R$ states (projection removes the cross terms $\bar{\Psi}_R\gamma^\mu\hat{\Psi}_L$ and $\bar{\Psi}_L\gamma^\mu\hat{\Psi}_R$). Hence we have
\be
j^\mu = \bar{\Psi}_L\gamma^\mu \hat{\Psi}_L + \bar{\Psi}_R\gamma^\mu \hat{\Psi}_R   \equiv j^\mu_L + j^\mu_R.
\label{jmu}
\ee

The sum of the currents in Eq. \ref{jj} leads to
\be
j^\mu + j^\mu_5 = \bar\Psi\gamma^\mu(1+\gamma_5)\hat{\Psi} = 2j^\mu_R,
\label{jplus}
\ee
whereas the difference yields
\be
j^\mu - j^\mu_5 = \bar\Psi\gamma^\mu(1-\gamma_5)\hat{\Psi} = 2j^\mu_L.
\label{jminus}
\ee
Equations \ref{jplus} and \ref{jminus} then reveal the simple relation
\be
j^\mu_5 = j^\mu_R - j^\mu_L.
\label{j5}
\ee
The axial current measures the difference of the charge and currents between the $L$ and $R$ populations; it is central to the chiral anomaly in Dirac/Weyl semimetals.

As mentioned, we have the two conservation laws
\begin{eqnarray}
\partial_{\mu}j^\mu &=& \frac{\partial \rho}{\partial t} + \nabla\cdot {\bf J} = 0, \label{Dj}\\
\partial_{\mu}j^\mu_5 &=& \frac{\partial \rho^5}{\partial t} + \nabla\cdot {\bf J}^5 = 0,
\label{Dj5}
\end{eqnarray}
with $j^\mu \equiv (\rho, {\bf J})^T$ and $j^\mu_5 \equiv (\rho^5, {\bf J}^5)^T$. 

These conservation laws are valid at the classical level. The process of quantization (brought about by coupling to vector gauge fields) destroys the conservation of $j^\mu_5$, with experimental consequences. This constitutes the chiral anomaly.

\subsection{Adler-Bell-Jackiw anomaly}\label{ABJ}
The chiral anomaly was discovered in the process of understanding conservation properties of the axial current and its role in pion decay~\cite{Adler,Bell,AdlerFifty,Jackiw}. The $\pi$ mesons $(\pi^+, \pi^0, \pi^-)^T$ are the lightest of all hadrons. Without any hadronic state to decay into, the primary channels are then leptonic (to electrons and muons); this results in a relatively long lifetime (26 ns) for the charged pions $\pi^{\pm}$. Remarkably, neutral pions $\pi^0$ decay 300 million times faster (lifetime 8.4$\times 10^{-17}$s) via the QED process $\pi^0\to 2 \gamma$ (this channel is forbidden for $\pi^{\pm}$ because of charge conservation). The struggle to calculate the amplitude ${\cal M}(\pi^0\to 2 \gamma)$ set the stage for the chiral anomaly. As pions are nearly ``massless'', the problem seemed tailor-made for the powerful current algebra techniques then in vogue, but these techniques yielded zero for ${\cal M}(\pi^0\to 2 \gamma)$ (the Veltman-Sutherland result). In 1968, Adler and Bell and Jackiw (ABJ) independently identified the correct process. The axial current is not conserved. Instead the conservation is violated by a source term ${\cal A}$, viz.
\be
\partial_{\mu}j^\mu_5 = {\cal A},
\label{DJ5}
\ee
with ${\cal A}$ (the Abelian anomaly term) given by
\be
{\cal A} = \frac{e^2}{16\pi^2} \varepsilon^{\mu\nu\alpha\beta}\; F_{\mu\nu}F_{\alpha\beta},
\label{A}
\ee
where $\varepsilon^{\mu\nu\alpha\beta}$ is the antisymmetric tensor and $F_{\mu\nu}$ the electromagnetic field tensor. The quantity ${\cal A}$ defined in Eq. \ref{A} is identical to the expression from charge pumping between chiral Landau levels (Eq. \ref{AA} in main text).

The lifetime of $\pi^0$ calculated from ${\cal A}$ achieved impressive agreement with experiment provided one assumed that quarks come in 3 colors (this was an early evidence for color). The ABJ triangle feynman diagram is the archetypal example for all subsequent anomaly calculations. Anomalies appear whenever a diagram has a fermion loop (quarks) coupled to vector currents (photons) and an odd number of axial currents ($\pi^0$).

In 1983, Nielsen and Ninomiya predicted~\cite{Nielsen} that the chiral anomaly should be observable as an unusual negative longitudinal magnetoresistance in semimetals that feature 3D Dirac states.

\subsection{Anomaly Cancellation}\label{cancel}
In the ABJ calculation of the decay rate of $\pi^0$, the triangle diagram linked an axial current (quarks) with 2 vector gauge bosons (photons). In electroweak theory and QCD, we can also have triangle diagrams with vertices linked only to gauge bosons. These diagrams are fatal because they render the theory unrenormalizable. Hence they must all cancel to zero. In the Glashow Weinberg Salam theory, each of the 3 vertices can be coupled to one of the 3 gauge bosons belonging to $U(1)$, $SU(2)$ or $SU(3)$. Peskin and Schroeder~\cite{Peskin} describe the cancellation of all the dangerous diagrams within each generation of quarks and leptons. The anomaly cancellation, dubbed magical~\cite{Peskin}, imposes constraints on the chirality of fermions ``running around'' the triangle loop.

\subsection{The $\mathbf{U_A(1)}$ problem}\label{U1}
Anomalies have played crucial roles in resolving deep puzzles at several energy hierarchies. By the early 70's quantum chromodynamics (QCD) based on quarks and leptons had gained universal acceptance, but a major hurdle remained -- the $U_A(1)$ problem~\cite{Peskin,ChengLi,Aitchison}. The small bare masses of the up and down quarks $u$ and $d$ (4 and 7 MeV, respectively) invite a massless fermion description. The invariance of ${\cal L}$ under the global axial unitary transformation $U_A(1)$ then suggests that the axial vector current is conserved, i.e. $\partial^\mu J^5_\mu = 0$. However, the assumed chiral symmetry immediately led to an unwelcome prediction: every hadronic state should have a parity pardner, in striking conflict with observation  (this is known as the parity-doublet problem). Hence the chiral symmetry must be spontaneously broken, but doing so raises a further problem -- too many Nambu-Goldstone bosons.

In a model incorporating just the lightest quarks $u$ and $d$, breaking of the chiral symmetry generates altogether 4 Nambu-Goldstone bosons -- the isotriplet which corresponds to the pions come from breaking of $SU_L\times SU_R$ symmetry plus an extra isoscalar meson. The last is conspicuously absent in the particle spectrum. When one includes the heavier strange quark $s$ (mass 130 MeV), a similar problem arises. There should be 2 isoscalar mesons, i.e. 9 low-mass mesons altogether but only 8 are observed (3 pions, 4 kaons and the $\eta$ particle). The pseudoscalar meson $\eta'$ is too massive to be part of the spectrum. This impasse constitutes the $U_A(1)$ problem~\cite{Peskin,ChengLi,Weinberg}.

To remove the unwanted $U_A(1)$ symmetry, one could invoke the anomaly mechanism, with ${\cal A}$ generalized to~\cite{ChengLi}
\be
{\cal A} = 4\frac{g^2}{16\pi^2} \mathrm{tr} (G_{\mu\nu}\tilde{G}^{\mu\nu}),
\label{GG}
\ee
where $g$ is the strong-interaction coupling parameter and tr means trace. The electromagnetic field tensor $F_{\mu\nu}$ in Eq. \ref{A} is replaced  by the gluon tensor matrix $G_{\mu\nu}$ (with $\tilde{G}^{\mu\nu}$ its dual). 
However, ${\cal A}$ is now a \emph{total} derivative, expressed as ${\cal A} = \partial^\mu K_\mu$, where $K_\mu$ acts as a current~\cite{ChengLi,Weinberg}. If we define a new axial vector current $\tilde{J}^5_\mu$ that includes $K_\mu$, viz. $\tilde{J}^5_\mu \equiv J^5_\mu - K_\mu$, we find that the corresponding axial charge $\tilde{Q}^5$ is conserved, i.e. 
\be
\partial^\mu \tilde{J}^5_\mu = 0.
\label{UA1}
\ee
Hence the $U_A(1)$ problem is apparently unresolved.

The solution discovered by `tHooft is that, in performing the path integrals, contributions from instantons must be included (see Sec. \ref{topology}).

\subsection{Topological Origin}\label{topology}
The ABJ anomaly was a bit of topology that fell into the physics of the 1960's. It soon emerged that $\cal A$ plays a role in gauge theory far more important than suggested by the resolution of the $\pi^0$ problem. An early hint to its topological origin was the finding that the one-loop triangle diagram receives no radiative corrections to all orders in perturbation calculations (the Adler-Bardeen theorem~\cite{AdlerBardeen}). Moreover, ${\cal A}$ is identical whether calculated from the triangle diagram (Eq. \ref{A}) or using Landau levels (Eq. \ref{AA}).

The central role of ${\cal A}$ may be seen from Fujikawa's path-integral approach to the anomaly~\cite{Fujikawa1979}. Under a transformation of the field variables in the Dirac fermion Lagrangian (hereafter, Euclidean metric is assumed), the Jacobian ${\cal J}$ acquires a phase determined by ${\cal A}$~\cite{Fujikawa,Bertlmann,Jackiw}. In particular, the axial rotation $\hat{\Psi}\rightarrow e^{i\beta\gamma^5}\hat{\Psi}$ induces the change ${\cal J}\rightarrow {\cal J}e^{-\beta\int d^4x\;{\cal A}(x)}$.

The phase in ${\cal J}$ involves the integral $\int d^4x\;{\cal A}(x)$ (the action ${\cal S}$) over a sphere in 4-space bounded by the 3-sphere $S_3$ of radius $R$. For ${\cal S}$ to be finite, the gauge potential $A_\mu$ must vanish rapidly as $R\to\infty$, to leave the gauge-only form $\hat{g}(\hat{x})^{-1}d\hat{g}(\hat{x})$, where $\hat{g}(\hat{x})$ is a transformation of the vacuum state in the direction $\hat{x}$ ($\hat{g}\in SU(2)$). Because $\hat{g}$ is specified by 3 parameters, just like the 3-sphere $S_3$, the mapping $S_3\to SU(2)$ partitions into homotopic sectors each characterized by an integer winding number $n\in \mathbb{Z}$ ($n$ is also called the topological charge or Pontryagin index)~\cite{ChengLi,Weinberg,Nakahara}. In Yang Mills theory, the winding number of the mapping arises from instantons (see below).

A lengthy calculation reveals that the integral $\int d^4x\;{\cal A}(x)$ equals an integer that measures the number of chiral zero modes (a zero mode $\phi_{0\pm}$ satisfies $\gamma^\mu D_\mu \phi_{0\pm} = 0$, where $D_\mu = \partial_\mu + A_\mu$ is the Dirac operator). We have
\be
\int d^4x \;{\cal A}(x)  = 2i(n_+ - n_-),
\label{intA}
\ee
where $n_{\pm}$ is the number of zero-modes of chirality $\pm$. The difference $(n_+ - n_-)$ is the index of the Weyl operator $D_+$~\cite{Bertlmann,Nakahara} (the index may be regarded as a generalization of the familiar Euler characteristic $\chi(M) = V-E+F$, an invariant that tracks the number of vertices $V$, edges $E$ and faces $F$ of a polyhedron~\cite{Nakahara}). 

Returning to the $U_A(1)$ problem, the apparent impasse, Eq. \ref{UA1}, is equivalent to assuming that the action $\int d^4x \;{\cal A}$ vanishes, i.e. the Jacobian ${\cal J}$ is unchanged. As shown by `tHooft, inclusion of instanton(s) makes the action finite instead (Eq. \ref{intA}). The instanton is a fluctuation of the vacuum in which $G_{\mu\nu}$ is finite within a local region of spacetime but dies away rapidly as $R\to \infty$. According to Weinberg~\cite{Weinberg}, the non-vanishing of the integral in Eq. \ref{intA} suffices to solve the $U_A(1)$ problem. The solution of the $U_A(1)$ problem provided an early hint of the rich structure of the QCD vacuum.

The calculation establishing Eq. \ref{intA} parallels that done for the Atiyah-Singer (AS) index theorem~\cite{Fujikawa1979,Bertlmann,Nakahara}. The AS index theorem is a far-reaching and deep generalization of the famous Gauss-Bonnet theorem, which relates the Gaussian curvature $K$ of a closed, compact and orientable surface $M$ to its Euler characteristic $\chi(M)$ by the integral
\be
\frac{1}{2\pi}\int_M \; KdA = \chi(M)= 2-2g_M,
\label{Gauss}
\ee
where the genus $g_M$ counts the number of holes in $M$~\cite{Nakahara,Frankel}. In Eq. \ref{intA}, ${\cal A}$ plays the role of $K$ while the index replaces $\chi(M)$. Thus, in addition to its central role in our picture of the QCD vacuum, the anomaly ${\cal A}$ is a central structure at the nexus of quantum physics and differential geometry.

\newpage


\newpage

\vspace{1cm}\noindent

\vspace{5mm}\noindent
{\bf Acknowledgements} We are indebted to Stephen Adler for valuable comments and suggestions.
N.P.O. acknowledges the support of the U.S. Army Research Office (ARO contract W911NF-16-1-0116), the U.S. National Science Foundation (Grant DMR 1420541) and the Gordon and Betty Moore Foundation’s EPiQS Initiative through Grant GBMF4539.

\newpage
\cleardoublepage

\end{document}